\documentclass[a4paper,11pt,amsmath,nofootinbib,superscriptaddress]{revtex4}
\usepackage{amsfonts}
\usepackage{graphicx}
\usepackage{comment}
\usepackage{array,amsmath,amsthm,feynmf}
\usepackage{bm}
\usepackage{slashed}
\usepackage{times}
\usepackage{epsfig}
\usepackage{graphicx}
\usepackage{color}
\usepackage{cancel}
\usepackage{amssymb}
\usepackage{textcomp}
\usepackage{mathrsfs}
\newcommand{\be}{\begin{equation}}
\newcommand{\ee}{\end{equation}}
\newcommand{\bea}{\begin{eqnarray}}
\newcommand{\eea}{\end{eqnarray}}
\def\shortsix{B_R^g}
\def\shortsixss{B_R^S}
\def\shortsixaa{B_R^{V}}
\def\shortS{N_S^{A}}
\def\shortSzw{N_S^{ZW}}
\def\namesix{R}
\def\nameS{S}
\def\namedm{A}
\def\swidthmin{10}
\def\swidthmax{50}
\def\gev{~\text{GeV}}
\def\fb{~\text{fb}}

\begin{document}



\title{Diphoton and Dark Matter from Cascade Decay }

\author{Fa Peng Huang}
\affiliation{Theoretical Physics Division, Institute of High Energy Physics,
Chinese Academy of Sciences, P.O.Box 918-4, Beijing 100049, P.R.China}

\author{Chong Sheng Li}
\affiliation{Department of Physics and State Key Laboratory of Nuclear Physics and Technology, Peking University, Beijing 100871, China}
\affiliation{Center for High Energy Physics, Peking University, Beijing 100871, China}


\author{Ze Long Liu}
\affiliation{Department of Physics and State Key Laboratory of Nuclear Physics and Technology, Peking University, Beijing 100871, China}

\author{Yan Wang
\footnote{wangyan728@ihep.ac.cn}}
\affiliation{Theoretical Physics Division, Institute of High Energy Physics,
Chinese Academy of Sciences, P.O.Box 918-4, Beijing 100049, P.R.China}

\date{\today}
\begin{abstract}
We propose a simplified model to study the possible new heavy diphoton resonance from cascade decay of a heavier particle at colliders, which may be related to the dark matter or other new physics beyond the standard model.
Model-independent constraints and predictions on the allowed couplings for generating the observed diphoton data are studied in detail.
We demonstrate that this scenario can be tested by the possible four-photon signal or the $WW/ZZ/Z\gamma$ resonances.
Meanwhile, this cascade decay scenario also provides us with the dark matter candidates, which is consistent with  the observed dark matter relic density.
\end{abstract}
\maketitle
\section{Introduction}\label{sec:Introduction}
New heavy resonances decaying into diphoton are predicted in many new physics models beyond the Standard Model
(SM).  And the heavy diphoton resonance can  provide a clean collider signature with accurate  invariant mass
resolution to search for new physics.
In this paper, we use the diphoton data recorded in 2015~~\cite{ATLAS-CONF-2015-081,ATLAS-CONF-2016-018,CMS-PAS-EXO-15-004,CMS-PAS-EXO-16-018}
as trial data to study the new heavy diphoton resonance and dark matter in the cascade decay scenario.
In 2015, both the ATLAS and CMS collaboration have released the consistent data on diphoton excess
around $750$ GeV at the 13 TeV LHC. The local statistical significance at ATLAS  (with $3.2~\text{fb}^{-1}$) is $3.9~\sigma$ and the local statistical significance at CMS (with $2.6~\text{fb}^{-1}$) is
$2.6~\sigma$.
The diphoton excess in the terms of cross section can be roughly estimated as~\cite{ATLAS-CONF-2015-081,ATLAS-CONF-2016-018,CMS-PAS-EXO-15-004,CMS-PAS-EXO-16-018}
\begin{eqnarray}\label{e_exp}
   \sigma_{\rm excess} &=& (10 \pm 3)~\text{fb} ~(\rm at~13~TeV ~ATLAS),\\
   \sigma_{\rm excess} &=& (6 \pm 3)~\text{fb}  ~~~(\rm at~13~TeV ~CMS).
\end{eqnarray}
This experiment immediately attracts extensive studies~\cite{Harigaya:2015ezk,Mambrini:2015wyu,Backovic:2015fnp,Angelescu:2015uiz,Nakai:2015ptz,Knapen:2015dap,Buttazzo:2015txu,Pilaftsis:2015ycr,Franceschini:2015kwy,DiChiara:2015vdm,Fiore:2015lnz,Higaki:2015jag,McDermott:2015sck,Low:2015qep,Bellazzini:2015nxw,Gupta:2015zzs,Petersson:2015mkr,Molinaro:2015cwg,Dutta:2015wqh,Cao:2015pto,McAllister:2015zcz,Matsuzaki:2015che,Kobakhidze:2015ldh,Martinez:2015kmn,Cox:2015ckc,Becirevic:2015fmu,No:2015bsn,Demidov:2015zqn,Chao:2015ttq,Fichet:2015vvy,Curtin:2015jcv,Bian:2015kjt,Chakrabortty:2015hff,Ahmed:2015uqt,Agrawal:2015dbf,Csaki:2015vek,Aloni:2015mxa,Gabrielli:2015dhk,Benbrik:2015fyz,Alves:2015jgx,Carpenter:2015ucu,Bernon:2015abk},
and the most direct explanation is that there exists a 750 GeV boson decaying into two photons, which is often constrained severely by the LHC Run-1 data.


However, another interesting diphoton excess at $M_{\gamma\gamma}\sim 1.6~ \text{TeV}$ is usually ignored. Its
local statistical significance is about $2.8$ $\sigma$~\cite{ATLAS-CONF-2015-081},  which is even higher than the $750$ GeV excess with $2.6~\sigma$
at the CMS~\cite{CMS-PAS-EXO-15-004}. Thus, there may also exist a 1.6 TeV boson, which may be produced via gluon fusion channel,
and can also decay to diphoton.

Incorporating both the 750 GeV and the 1.6 TeV diphoton excess, we propose a natural
interpretation by assuming that a much more heavier particle, which is produced by gluon fusion, decays to the 750 GeV boson.
Then the 750 GeV boson sequentially decays to diphoton \footnote{Although the new bosons can be $\rm spin$-$0$ or $\rm spin$-$2$, we only investigate
the scalar  boson resonances in this paper for simlicity.}, but it can not interact with gluon and quarks.
Therefore, in this cascade decay scenario, the  constraints from the 8 TeV LHC can be easily avoided,
since the  production rate of the heavier particle is highly
suppressed by the relative low center-of-mass (c.m.) energy.
As for the decay channels of the 750 GeV boson, we firstly investigate the most
simplest case: the 750 GeV boson can only couple to the photon or an invisible particle. And the dominant decay channel is the invisible decay, which can naturally provide the dark matter (DM) candidate;
the second decay channel is the diphoton channel.
As a result, the diphoton excess can be explained by the $\gamma\gamma$ + missing transverse energy (MET) production, where the MET is too small to be observed in the experiment by choosing appropriate parameters. This scenario also predicts the existence of four-photon signals at high
luminosity LHC and the 1.6 TeV diphoton signals,
which  can be tested in the future. For completeness, we also  discuss the interactions of the 750 GeV new boson with the $Z$ boson
and $W$ boson. The allowed parameter spaces and the predicted signals
are given for different cases. After the new data recorded in 2016 are released, we
update the discussions.

This paper is organized as follows. In section \ref{sec:Model}, we present the  effective Lagrangian and perform the
general discussions, which can explain the 750 GeV diphoton excess
and the 1.6 TeV  diphoton signatures.
In section \ref{sec:CaseI}, the detailed discussions for the case without W or Z boson  is given.
In section \ref{sec:CaseII}, the cases including W or Z boson are also discussed for integrity.
Finally, we update the discussions based on the new experimental data and conclude in section \ref{sec:conclusion}.

\section{The effective Lagrangian}\label{sec:Model}
Instead of investigating the diphoton excess in a concrete UV-complete theory
(such as some  modified composite Higgs model), which is not easy to make clear
predictions originating from large sets of undetermined
model parameters, we take a bottom-up approach to explain
the diphoton excess using the effective field theory.
Thus,
we begin our investigation on the diphoton excess from the following effective Lagrangian:

\begin{eqnarray}
\label{eq:lang}
\mathcal{L}_{\text{eff}}&=&\frac{c_{\namesix gg} }{\Lambda_1} G^b_{\mu\nu}G^{b\mu\nu} \namesix   +  \frac{c_{\namesix B}}{\Lambda_2} B_{\mu\nu}B^{\mu\nu} \namesix + \frac{c_{\namesix W}}{\Lambda_2} W_{\mu\nu}W^{\mu\nu} \namesix + c_{\namesix  \nameS \nameS} \Lambda_3 \namesix \nameS \nameS \nonumber\\
&+& \frac{c_{\nameS B}}{\Lambda_4} B_{\mu\nu}B^{\mu\nu} \nameS + \frac{c_{\nameS W}}{\Lambda_4} W_{\mu\nu}W^{\mu\nu} \nameS +c_{ \nameS \namedm \namedm}  \Lambda_5 \namedm \namedm \nameS.
\end{eqnarray}
where $B_{\mu\nu}, W_{\mu\nu}$ and $G_{\mu\nu}^{b}$ represent the SM $U(1)_Y$, $SU(2)_L$ and $SU(3)_c$ field strength tensor, respectively.
$\nameS$ is a scalar field with its mass $M_{\nameS}=750$~GeV.
$\namesix$ is another very heavy scalar field, which can decay to $\nameS$. If the mass of $\namesix$ is  much heavier than $750$ GeV, the constraints from
the $8$ TeV LHC data~\cite{Khachatryan:2015qba} can be naturally evaded.
We set the default mass of the particle $\namesix$ as  $M_{\namesix}=1.6$~TeV, which corresponds to the second rensonance in ATLAS experiment~\cite{ATLAS-CONF-2015-081}. Since the information of  this excess is little now, we assume the width of $\namesix$ to be less than 100 GeV.
Particle $\namedm$ is an invisible stable scalar field, which can be recognized as a DM candidate.
In general, both $\namesix$ and $\nameS$ can interact with $B_{\mu\nu}$ and $W_{\mu\nu}$. After electroweak symmetry breaking, $\namesix$ and $\nameS$ can couple with photon, $Z$ and $W$ bosons, that is why there exists diphoton excess at $750 \gev$ and $1600 \gev$.
$c_{\namesix gg},c_{\namesix \nameS \nameS},c_{\namesix B},c_{\namesix W},c_{\nameS B},c_{\nameS W}$ and $c_{ \nameS \namedm \namedm}$ are the dimensionless couplings and
$\Lambda_1,\Lambda_2,\Lambda_3,\Lambda_4,\Lambda_5,\Lambda_6$ are the dimension-1 energy scales.
For simplicity, in our following analysis and calculations, we set
 $\Lambda_{i}=10^4\gev~(i=1,...,6)$.

These effective operators can be obtained in some concrete
renormalizable models, such as the extended SM with
triplet Higgs and the vector-like quarks~\cite{Huang:2015izx} or the
extended Weinberg Higgs portal model with excited quarks~\cite{Weinberg:2013kea,Huang:2013oua,Zhan:2015dha}.
For example, the dimension-5 effective operator $ \frac{c_{\namesix gg}}{\Lambda_1} G^b_{\mu\nu}G^{b\mu\nu} \namesix $ in Eq.~(\ref{eq:lang}) can be realized from a renormalizable
perturbative theory via loops of several heavy colored
vector-like fermions with heavy mass, which can easily avoid  the
constraints from the current Higgs data.
For the simplest case, we can add the color triplet vector-like fermions $\chi$ and $\chi^c$
by the following interaction with the heavy scalar particle $\namesix$
\begin{equation}
\delta \mathcal{L}=-y_{\namesix} \chi \chi^c  \namesix.
\end{equation}
By integrating out the heavy fermion in the loop ,
the effective dimension-5 interaction between the heavy
scalar field $\namesix$ and two gluons can be obtained as
\begin{equation}
\delta \mathcal{L}_{eff}=\frac{\alpha_s y_{\namesix} f_{\chi}}{12 \pi M_{\chi}} G^b_{\mu\nu}G^{b\mu\nu}  \namesix,
\end{equation}
where $f_{\chi}$ is the loop integrated function
\begin{equation}
f_{\chi}=\frac{3}{2 x^2} \big(x+(x-1)f(x)\big),
\end{equation}
in which $x \equiv (M_{\namesix}/(2M_{\chi}))^2$, and
\begin{eqnarray}
  f(x) &=& \arcsin^2 \sqrt{x} ~~~~~~~~ \text{for}~ x \leq 1 , \nonumber\\
  f(x) &=& -\frac{1}{4} (\log\frac{1+\sqrt{1-x^{-1}}}{1-\sqrt{1-x^{-1}}}-i \pi)^2  ~~~~~~~~ \text{for}~~ x >1.
\end{eqnarray}
Matching to the effective Lagrangian in Eq.~(\ref{eq:lang}),
the matching coefficient can be given as
\begin{equation}
\frac{c_{\namesix gg}}{\Lambda_1}=\frac{\alpha_s y_{\namesix}}{12 \pi M_{\chi}} f_{\chi}.
\end{equation}
By the similar procedure, the other dimension-5 effective operators can  be induced by introducing charged
heavy fermions on in the loop.

After the electroweak symmetry breaking, the scalar $\namesix$ and $\nameS$ can both decay to $\gamma\gamma$, $WW$, $ZZ$ and $Z\gamma$. We list the decay widths as follows
\begin{eqnarray}
\Gamma(\namesix\to g g)&=&  \frac{2 c_{\namesix gg}^2 M_{\namesix}^3 }{\pi \Lambda_1^2}, \nonumber \\
\Gamma(\namesix \to \nameS \nameS)&=& \frac{c_{r\nameS\nameS}^2 \Lambda_3^2}{8\pi M_{\namesix}}\sqrt{1-\frac{4 M_{\nameS}^2}{M_{\namesix}^2}},\nonumber\\
\Gamma(\nameS \to \namedm \namedm)&=&\frac{c_{\nameS a a}^2 \Lambda_5^2}{8\pi M_{\nameS}}\sqrt{1-\frac{4 M_{\namedm}^2}{M_{\nameS}^2}},\nonumber\\
\Gamma(X \to \gamma\gamma)&=& \frac{M_{X}^3 }{4 \pi \Lambda_X^2}(c_{X W} s_w^2 + c_{X B} c_w^2)^2,\nonumber\\
\Gamma(X \to Z\gamma) &=& \frac{M_{X}^3 }{2 \pi \Lambda_X^2}(c_{X W} - c_{X B})^2 (1-\frac{M_Z^2}{M_{X}^2})^3 c_w^2 s_w^2,\nonumber\\
\Gamma(X \to WW)&=& \frac{M_{X}^3 }{2 \pi \Lambda_X^2}(1-6\frac{M_W^2}{M_{X}^2})c_{X W}^2,\nonumber\\
\Gamma(X \to ZZ)&=& \frac{M_{X}^3 }{4 \pi \Lambda_X^2}(c_{X W} c_w^2 + c_{X B} s_w^2)^2 (1-6\frac{M_Z^2}{M_{X}^2}),
\end{eqnarray}
where $X=\namesix$ or $\nameS$, $\Lambda_X=\Lambda_2$ or $\Lambda_4$, $s_w\equiv sin \theta_w$,~$c_w\equiv cos\theta_w$,  and $M_{Z/W}$ is the mass of the $Z/W$ boson. We define the effective coupling of S particle with the $\gamma\gamma$, $Z\gamma$, $ZZ$ and $WW$ as $c_{S\gamma\gamma}$, $c_{SZ\gamma}$, $c_{SZZ}$ and
$c_{SWW}$, respectively. Their expressions in terms of the original couplings can be obtained from the above formulae of the decay widths.

\begin{figure}
  \centering
  \includegraphics[width=.33\textwidth]{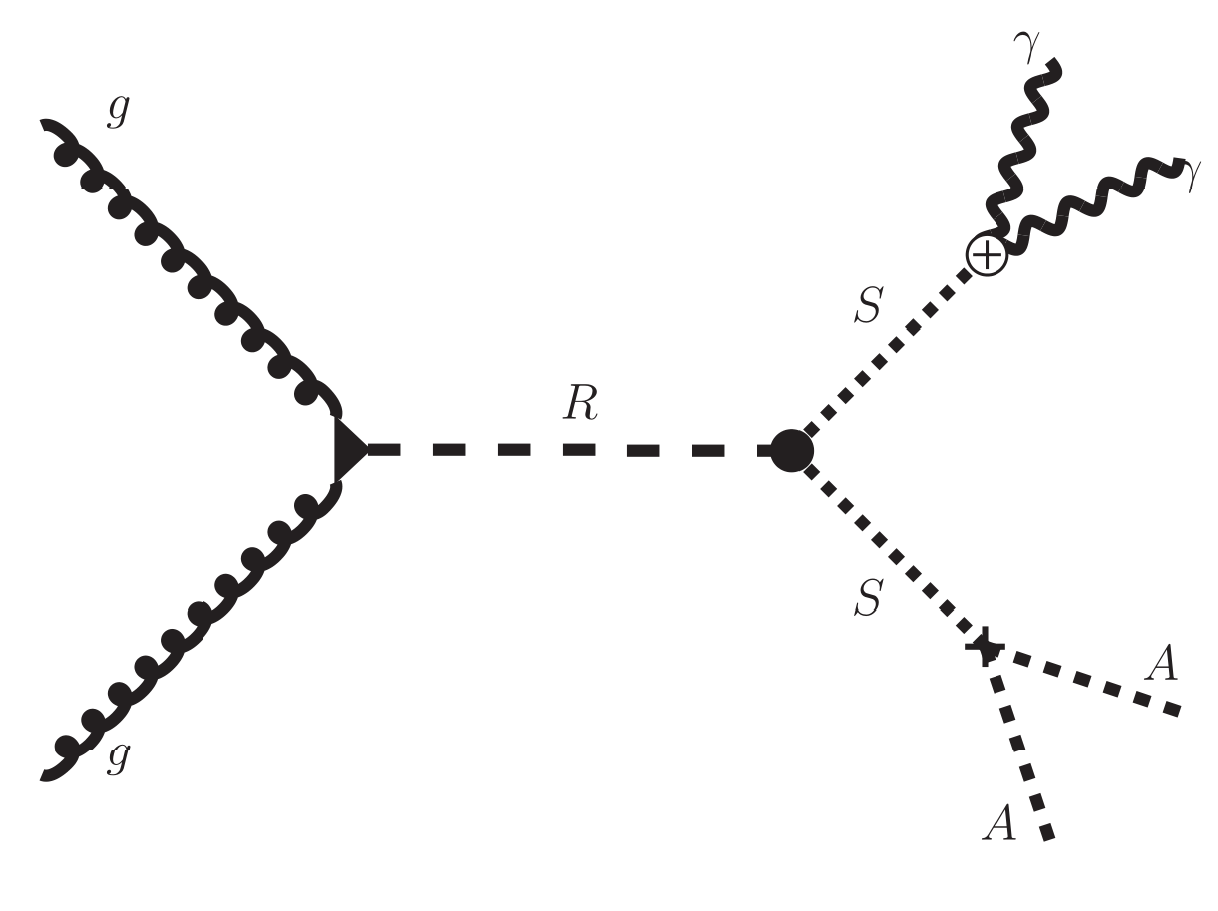}\\
  \caption{The Feynman diagram for the production of the 750 GeV diphoton excess.}\label{production}
\end{figure}
As a result, the 750 $\gev$ diphoton excess can be explained easily from the following
cascade decay process:
\begin{equation}\label{ssig}
\sigma_{\text{excess}}(pp\to \namesix \to \nameS \nameS \to \gamma \gamma \namedm \namedm) = \sigma(pp\to \namesix \to \nameS \nameS) Br(\nameS \to \namedm \namedm) Br(\nameS \to \gamma \gamma)\approx10~\rm fb.
\end{equation}
The corresponding Feynman diagram is shown in Fig.~\ref{production}.

Because the transverse momentum of the scalar $\namesix$ vanishes at leading order, and when $M_{\namesix}\sim 2~M_{\nameS}$, the boost of $\nameS$ is heavily suppressed.
Thus, the total transverse momentum of the decay product of S ($\namedm$ pair or diphoton) are small.
Since $\namedm$ pair is invisible and it can only be resolved at large MET,  the process in Eq.(\ref{ssig}) would be recognized as diphoton production when $M_{\namesix}\to2M_{\nameS}$.
The MET is about $E_T^{missing}\sim \sqrt{M_{\nameS}(M_{\namesix}-2M_{\nameS})}$, which means that MET highly depends on the difference between $M_{\namesix}$ and $2 M_{\nameS}$, and its distribution will be discussed in details later.

\begin{figure}
  \centering
  \includegraphics[width=.33\textwidth]{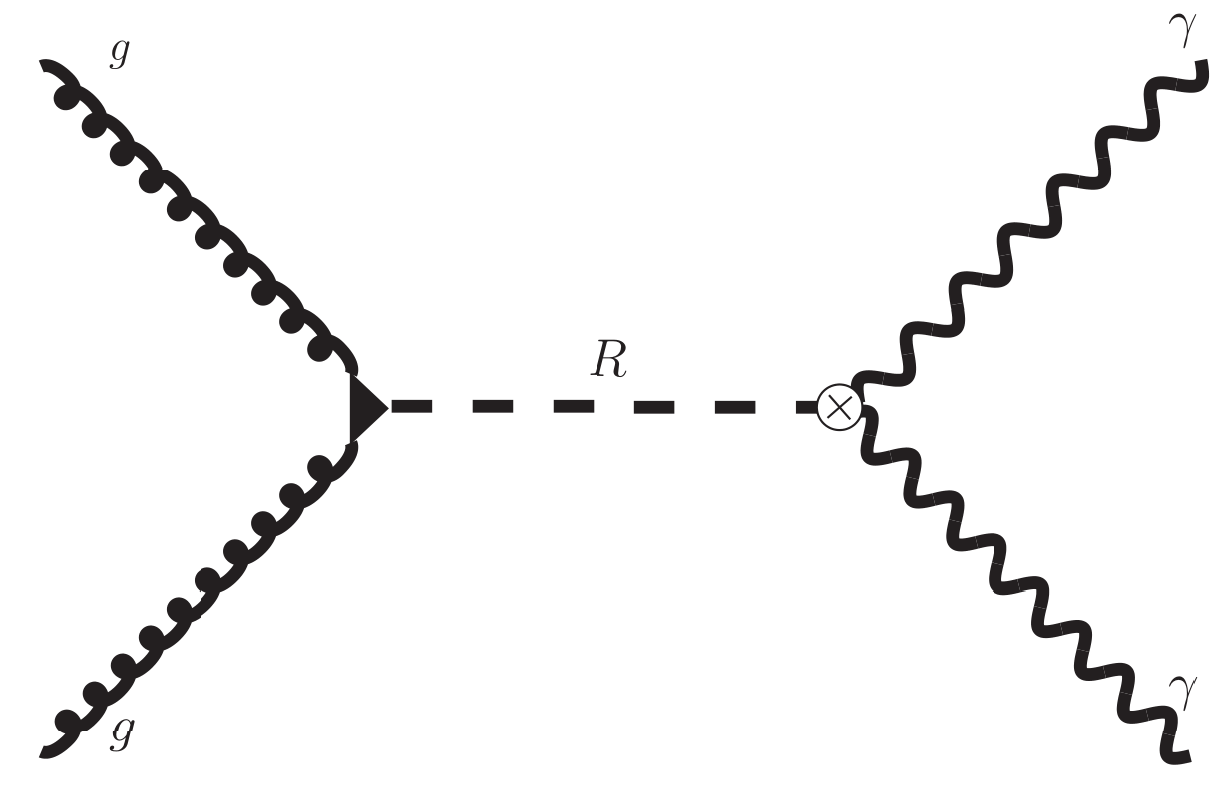}\\
  \caption{The Feynman diagram for the production of the 1.6 TeV diphoton excess.}\label{f_fm_2gamma}
\end{figure}

Another interesting deviation concerned here is at $M_{\gamma\gamma}\sim 1.6~ \text{TeV}$, where the cross section is about 1 $fb$ and the local
statistical significance is about $2.8$ $\sigma$~\cite{ATLAS-CONF-2015-081}. It can be explained  as the process $gg \rightarrow R \rightarrow \gamma\gamma$,
and the Feynman diagram is shown in Fig.~\ref{f_fm_2gamma}. This process will affect the total decay width of $\namesix$.

For convenience, we define the ratios of decay widths
\begin{eqnarray}
  \shortsix &=&  \frac{\Gamma(\namesix \to g g)}{\Gamma(\namesix)}, \nonumber\\
  \shortsixss &=&  \frac{\Gamma(\namesix \to \nameS\nameS)}{\Gamma(\namesix)}, \nonumber\\
  \shortsixaa &=&  \frac{\Gamma(\namesix \to \gamma\gamma)+\Gamma(\namesix \to Z\gamma)+\Gamma(\namesix \to ZZ)+\Gamma(\namesix \to WW)}{\Gamma(\namesix)}, \nonumber\\
  \shortS &=& \frac{\Gamma(\nameS \to \namedm  \namedm)}{\Gamma(\nameS \to \gamma \gamma)},\nonumber\\
  \shortSzw &=& \frac{\Gamma(\nameS \to ZZ)+\Gamma(\nameS \to Z\gamma)+\Gamma(\nameS \to WW)}{\Gamma(\nameS \to \gamma \gamma)},
\end{eqnarray}
Then, Eq.~(\ref{ssig}) can be written as
\begin{equation}\label{sss}
 \sigma_{\text{excess}}=\frac{f_{gg}}{M_{\namesix}  S_{c.m.}} \Gamma(\namesix)\shortsix \shortsixss  \frac{2 \shortS}{(\shortS+\shortSzw+1)^2},
\end{equation}
where the c.m. energy $ \sqrt{S_{c.m.}}=13~\rm TeV$, and $f_{gg}$ is defined as
\begin{equation}
f_{gg}=\frac{\pi^2}{8}\int ^1_{M_{\namesix}^2/S_{c.m.}} \frac{dx}{x}f_g(x)f_g(\frac{M_{\namesix}^2}{S_{c.m.}x}),
\end{equation}
wherein $f_g(x)$ is the gluon parton distribution function.
Here, we use the MSTW2008NLO~\cite{Martin:2009iq} to perform numerical calculations
and
obtain the concise result
\begin{equation}\label{nums}
\sigma_{\text{excess}}=89.36~\frac{ \Gamma(\namesix)}{\rm GeV}  \frac{\shortsix\shortsixss \shortS}{(\shortS+\shortSzw+1)^2} \rm fb.
\end{equation}

Since the decay channel $\namesix \to g g$ exists, extra dijet events must be suppressed to satisfy the current experimental data.
We set a stringent upper bound on the cross section that $\sigma(pp \to \namesix \to gg)  < 100$~fb, which results in
$0.1 \leq \shortsix \leq \frac{1}{2}$, and it can be easily consistent with bounds of the dijet resonance searching data~\cite{CMS-PAS-EXO-15-001}.  We also require $\Gamma_{\namesix}< 100 \rm ~GeV$.

To avoid large
four-photon cross section, the following condition is necessary,
\begin{equation}\label{c1}
\Gamma(\nameS \to \gamma\gamma)< \Gamma(\nameS \to \namedm \namedm),
\end{equation}
namely, $\shortS\geq 1$.
Thus, the four-photon signal is weak.  And we also suppose $\shortS < 10$ in the following discussing, and the four-photon signal may be detected at high luminosity LHC.
Notice that when $\shortSzw$ is fixed, $\frac{\shortS}{(\shortS+\shortSzw+1)^2}$ is a  monotonic decreasing function on $\shortS$,
when $\shortS\geq 1$ and when $\shortSzw$ is small, we always have
$\frac{\shortS}{(\shortS+1)^2} \leq \frac{1}{4}$,
namely,
\begin{equation}
Br(\nameS \to \gamma \gamma) \times  Br(\nameS \to AA) \leq \frac{1}{4}.
\end{equation}
The value of $c_{\nameS\gamma\gamma}$ depends on the width of the particle $\nameS$ and $\shortS$, and we assume that $\swidthmin\gev <\Gamma_{\nameS} < \swidthmax\gev$.

In order to show the features of different parameter setup, and maximize the branch ratio of $\nameS \rightarrow \gamma\gamma$, we compare the following three cases:
\begin{enumerate}
  \item Case I: $\nameS$ can only  decay to $\gamma\gamma$ and $\namedm\namedm$. As a result, $\shortSzw$=0 and there are only three decay channel for $\nameS$: $\namesix \to (\nameS \to \namedm \namedm) (\nameS \to \namedm \namedm)$, $\namesix \to (\nameS \to \namedm \namedm)(\nameS \to \gamma \gamma)$ and $\namesix \to (\nameS \to \gamma \gamma)(\nameS \to \gamma \gamma)$. In this case,  $c_{S\gamma\gamma}\neq 0$, $c_{SZ\gamma}=c_{SWW}=c_{SZZ}=0$.
  \item Case II: Set $c_{\nameS W}=0$ and ignore the interaction between $R$ and $\gamma/Z/W$. As a result, $\Gamma(\nameS\rightarrow WW)=0$, $\Gamma(\nameS\rightarrow ZZ)\sim0.08~\Gamma(\nameS\rightarrow \gamma\gamma)$ and $\Gamma(\nameS\rightarrow \gamma Z) \sim 0.58~\Gamma(\nameS\rightarrow \gamma\gamma)$, which means $\shortSzw$=0.66. Namely, $c_{SWW}=0$, $c_{SZ\gamma}\approx 0.3 ~c_{S\gamma\gamma}$ and $c_{SZ\gamma}\approx 0.78 ~c_{S\gamma\gamma}$.
  \item Case III: Set  $c_{\nameS W}=c_{\nameS B}$ and ignore the interaction between $R$ and $\gamma/Z/W$. In this case, $\Gamma(\nameS\rightarrow WW) \sim 1.9~\Gamma(\nameS\rightarrow \gamma\gamma)$ and $\Gamma(\nameS\rightarrow ZZ) \sim 0.9~\Gamma(\nameS\rightarrow \gamma\gamma)$, as well as $\Gamma(\nameS\rightarrow Z\gamma)=0$. Therefore, $\shortSzw\sim2.8$, and the production of $WW$ or $ZZ$ events will become equally important than diphoton in the experiment, which will be constrained  by the diboson resonance data.
      Namely, $c_{SZ\gamma}=0$, $c_{SZZ}\approx 0.95~c_{S\gamma\gamma}$ and $c_{SWW}\approx 1.38~c_{S\gamma\gamma}$.
\end{enumerate}
Here, Case I is an artificial example, which ignores the possible interactions  $SZZ$,  $SWW$ and $SZ\gamma$.
In principle, from the Lagrangian given in Eq.(\ref{eq:lang}),
it is impossible to obtain Case I directly.
Case I  is shown for clearly
explaining the cascade decay scenario, which
can explain the diphoton excess and the DM.
Further, this simple case can simplify the discussion
and provide us with appropriate benchmark parameter sets.
These benchmark sets help to optimize the discussion process in
Case II and Case III, which can be obtained from the lagrangian in
Eq.(\ref{eq:lang}).

\section{parameter spaces in  Case I}\label{sec:CaseI}
In Case I, to be compatible with the current 8 TeV data and the 13 TeV data at the LHC, we assume that the dominant decay channel of $\nameS$ is the
invisible decay $\namesix \to (\nameS \to \namedm \namedm) (\nameS \to \namedm \namedm)$,   the second decay channel is just the observed
$750$~GeV diphoton excess $\namesix \to (\nameS \to \namedm \namedm)(\nameS \to \gamma \gamma)$, and the smallest decay mode is the four-photon decay
channel $\namesix \to (\nameS \to \gamma \gamma)(\nameS \to \gamma \gamma)$.

In order to simplify the discussions, we select several concrete benchmark sets to discuss the diphoton excess at the 13 TeV LHC.
\begin{eqnarray}
      && \text{benchmark}~ 1:~~~~~~\shortsix = \frac{1}{2} ,~~~~~~\shortS = 1, ~~~~~~~~ \text{benchmark}~ 2:~~~~~~\shortsix = \frac{1}{2} ,~~~~~~\shortS = 10,\nonumber\\
      && \text{benchmark}~ 3:~~~~~~\shortsix = \frac{1}{11},~~~~\shortS = 1, ~~~~~~~~ \text{benchmark}~ 4:~~~~~~\shortsix = \frac{1}{11},~~~~\shortS = 10,\nonumber   \\
      && \text{benchmark}~ 5:~~~~~~\shortsix = \frac{1}{9} ,~~~~~~\shortS = 5.
\end{eqnarray}

\subsection{diphoton excess at 750 GeV}
\begin{figure}[h]
\centering
\begin{minipage}[t]{0.33\linewidth}
\centering
 \includegraphics[width=1.0\linewidth]{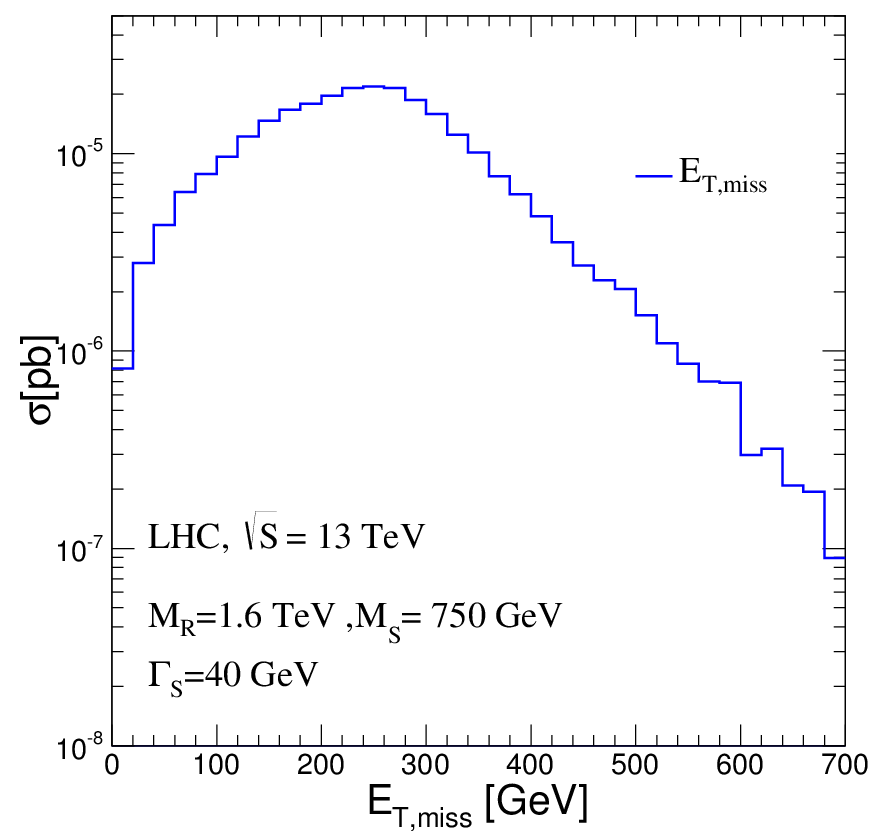}\\
   \text{(a)}
\end{minipage}
\hfill
\begin{minipage}[t]{0.33\linewidth}
\centering
 \includegraphics[width=1.0\linewidth]{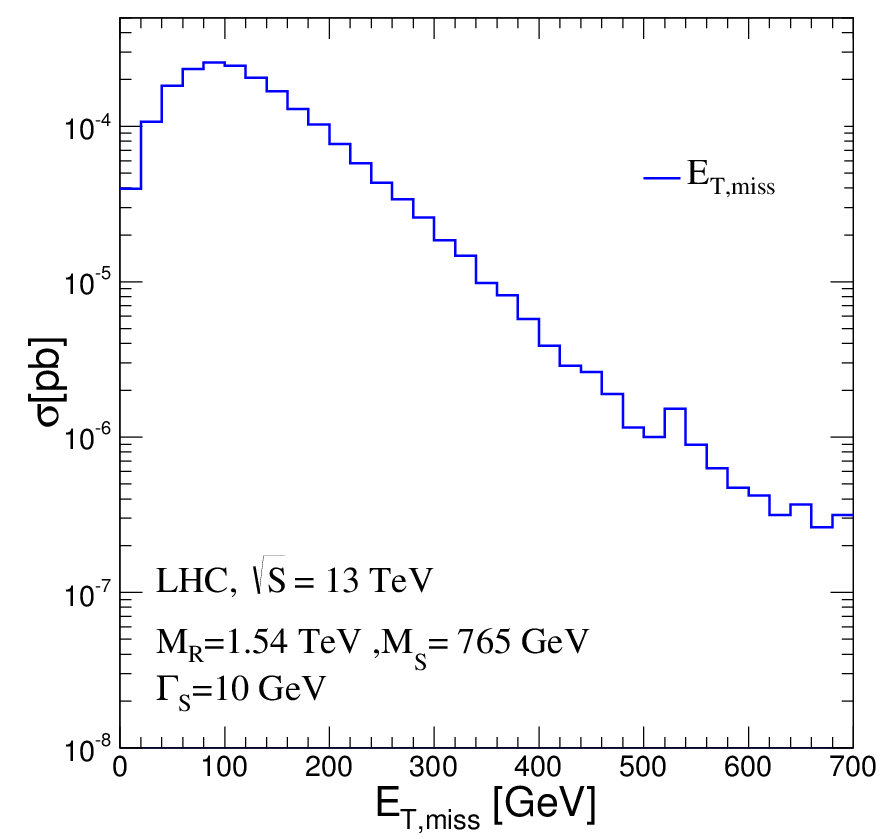}\\
   \text{(b)}
\end{minipage}
\hfill
\begin{minipage}[t]{0.33\linewidth}
\centering
 \includegraphics[width=1.0\linewidth]{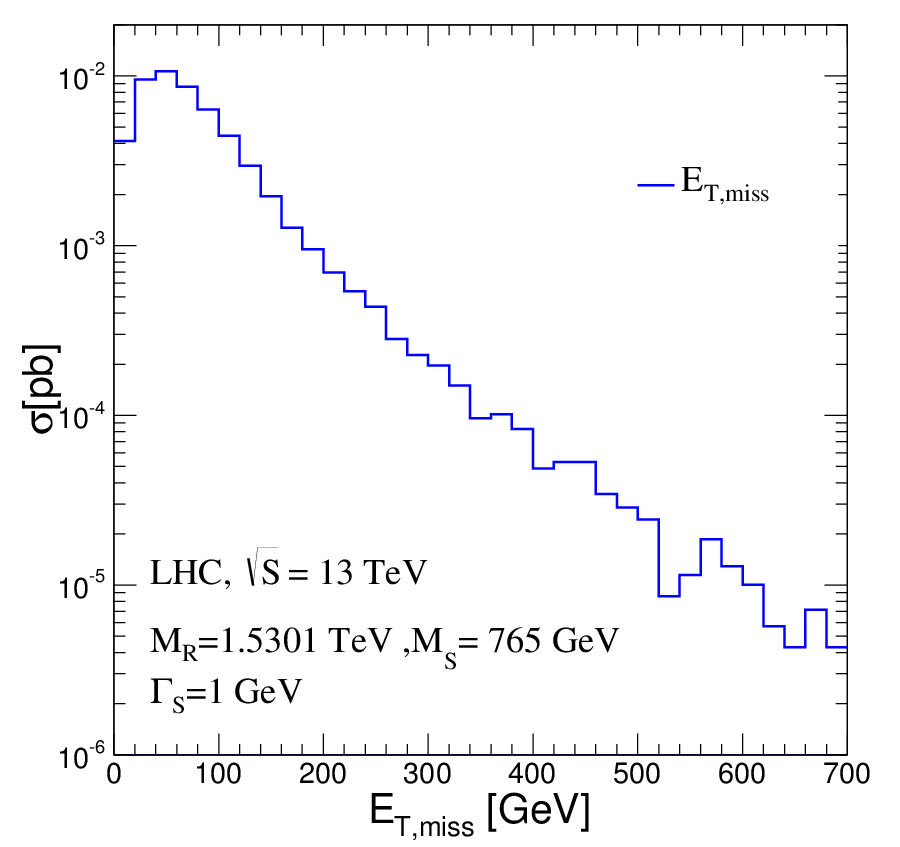}\\
   \text{(c)}
\end{minipage}
\caption{The MET distribution in $\namesix \to (\nameS \to \namedm \namedm)(\nameS \to \gamma \gamma)$ channel. The three figures correspond to (a) $M_{\namesix}=1600\gev, M_{\nameS}=750\gev ~\text{and}~ \Gamma(\nameS)=50\gev$; (b) $M_{\namesix}=1540\gev, M_{\nameS}=765\gev ~\text{and}~ \Gamma(\nameS)=10\gev$ and (c) $M_{\namesix}=1530.1\gev, M_{\nameS}=765\gev ~\text{and}~  \Gamma(\nameS)=1\gev$.
}\label{f_met}
\end{figure}

In order to generate the observed diphoton excess, it requires that the MET is small enough to be ignored in the experiment. In Fig.~\ref{f_met}, we show the MET distribution in the $gg\to \gamma\gamma AA$ channel, where the $\gamma\gamma AA+0/1/2~\text{jets}$ parton-level matched samples are generated with MadGraph5~\cite{Alwall:2014hca} at leading order. We choose three parameter sets, which are within the uncertainties of the experiment: (a) $M_{\namesix}=1600\gev, M_{\nameS}=750\gev ~\text{and}~ \Gamma(\nameS)=50\gev$; (b) $M_{\namesix}=1540\gev, M_{\nameS}=765\gev ~\text{and}~ \Gamma(\nameS)=10\gev$ and (c) $M_{\namesix}=1530.1\gev, M_{\nameS}=765\gev ~\text{and}~ \Gamma(\nameS)=1\gev$. It is obvious that the MET highly depends on the mass as well as width of $\namesix$ and $\nameS$. The peaks of the MET distribution in three figures are at about $E_T^{missing}\sim \sqrt{M_{\nameS}(M_{\namesix}-2M_{\nameS})}$. As a result, if the mass difference between $M_{\namesix}$ and $2M_{\nameS}$ is small,
the $gg\to \gamma\gamma AA$ channel can be identified as the diphoton excess at the 13 TeV LHC.

\begin{figure}[h]
\centering
\includegraphics[width=0.45\linewidth]{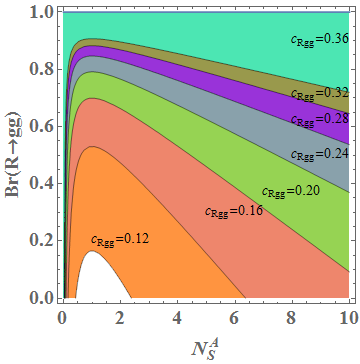}\\
\caption{The constraints for $\shortS$ and $\shortsix$ when fixing $c_{\namesix gg}$. In order to generate diphoton excess with $10\pm3$ fb cross sectioin, $c_{\namesix gg}$ need to be between 0.2-0.5 for different benchmark points. The colored bands stand for the cross section of diphoton excess which is smaller than 7 fb when fixing different $c_{\namesix gg}$, respectively.
}\label{f_nganr}
\end{figure}

As shown in Eq.(\ref{sss}), the diphoton excess is related to four parameters, i.e., $c_{Rgg}$, $\shortsix$, $\shortsixss$ and $\shortS$.
Figure~\ref{f_nganr} presents the diphoton cross sections in $\shortS$-$\shortsix$ plane, where we choose $c_{\namesix gg}=0.1-0.4$ and fix  $\shortsixss=1-\shortsix$ (i.e. $\shortsixaa=0$, we will show the rationality in the next subsection), respectively.
The colored region corresponds to the signal cross section larger than 7 fb. It is obviously that there exists large parameter spaces for the observed diphoton excess.

\begin{figure}[h]
\centering
\includegraphics[width=0.45\linewidth]{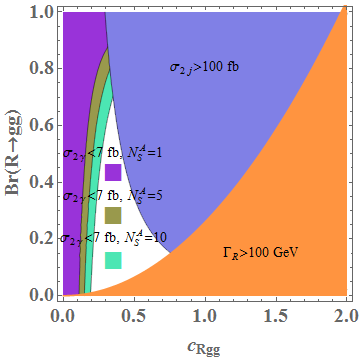}\\
\caption{The contour plot for $c_{\namesix gg}$ and  $\shortsix$, assuming diphoton excess larger than 7 fb. The purple, wheat and green regions denote that diphoton excess is small than 7 fb, when fixing $\shortS=1,5,10$, respectively. The orange region stands for $\Gamma_{\namesix}>100~\text{GeV}$, and the blue region means $\sigma_{dijet}>100 fb$. The white region stands for the allowed parameter space.
}\label{f_cpggandnp}
\end{figure}

The main constraints on the $c_{\namesix gg}$-$\shortsix$ plane is shown in Fig.\ref{f_cpggandnp}. Firstly, since we assume that the width of $\namesix$ is smaller than 100~GeV, it can only provide a loose constraint on $c_{\namesix gg}$ and $\shortsix$, denoted by the orange color region.  Secondly, the blue region stands for the parameter spaces with $\sigma(gg \rightarrow \namesix \rightarrow j j)>100~\text{fb}$. Finally, in the parameter spaces that we concerned, the  most strict constraint comes from that the diphoton excess need to be large than 7 fb, which is the minimum value of ATLAS experiment in Eq.~(\ref{e_exp}).   The parameter spaces in the purple (wheat and green) region denotes that $\sigma_{excess}(gg\rightarrow 2\gamma2A) < 7~\text{fb}$, where $\shortS$ is fixed at $1$ ($5$ and $10$). Finally, the white region stands for the allowed parameter space for diphoton excess.
After combining those constraints, we also list the main constraints for the benchmark sets in Table~\ref{constrainforB}.
\begin{table}[h]
\begin{center}
\begin{tabular}{|c|c|c|c|c|c|}
  \hline
  benchmark sets & 1 & 2 & 3 & 4 & 5 \\
  \hline
  $c_{\namesix gg}$ & [0.16,~0.21] & [0.27,~0.37] & [0.11,~0.16] & [0.20,~0.27] & [0.16,~0.21] \\
  \hline
  $\Gamma_{\namesix} (\text{GeV})$ & [1.3,~2.3] & [3.8,~7.1] & [3.5,~7.3] & [11.5,~20.9] & [6.0,~10.3] \\
  \hline
  $c_{\nameS\gamma\gamma}$ & [6.68,~8.63] & [2.85,~3.68] & [6.68,~8.63] & [2.85,~3.68] & [3.85,~4.98] \\
  \hline
\end{tabular}
\end{center}
  \caption{The combined constraints for  $c_{\namesix gg}$, $\Gamma_{\namesix}$  and  $c_{\nameS\gamma\gamma}$  on each benchmark point. The results come from combining the constraints from the $\swidthmin\gev<\Gamma_{\nameS}<\swidthmax\gev$,
$\Gamma_{\namesix}<100~\text{GeV}$, the $10\pm3$ fb diphoton excess and the dijet resonances searching.}\label{constrainforB}
\end{table}

\subsection{1.6 TeV diphoton resonance predictions}
There is another interesting deviation from the background around $M_{\gamma\gamma}\sim 1.6~ \text{TeV}$ in the ATLAS experiment, where the cross section is smaller than $1\fb$ and the local
statistical significance is about $2.8$ $\sigma$~\cite{ATLAS-CONF-2015-081}. As we stated above, this deviation can be explained naturally by the scalar $\namesix$ with mass 1.6TeV.
The Feynman diagram for diphoton production induced by $\namesix$ is shown in Fig.~\ref{f_fm_2gamma}.

\begin{figure}[h]
\centering
 \includegraphics[width=0.45\linewidth]{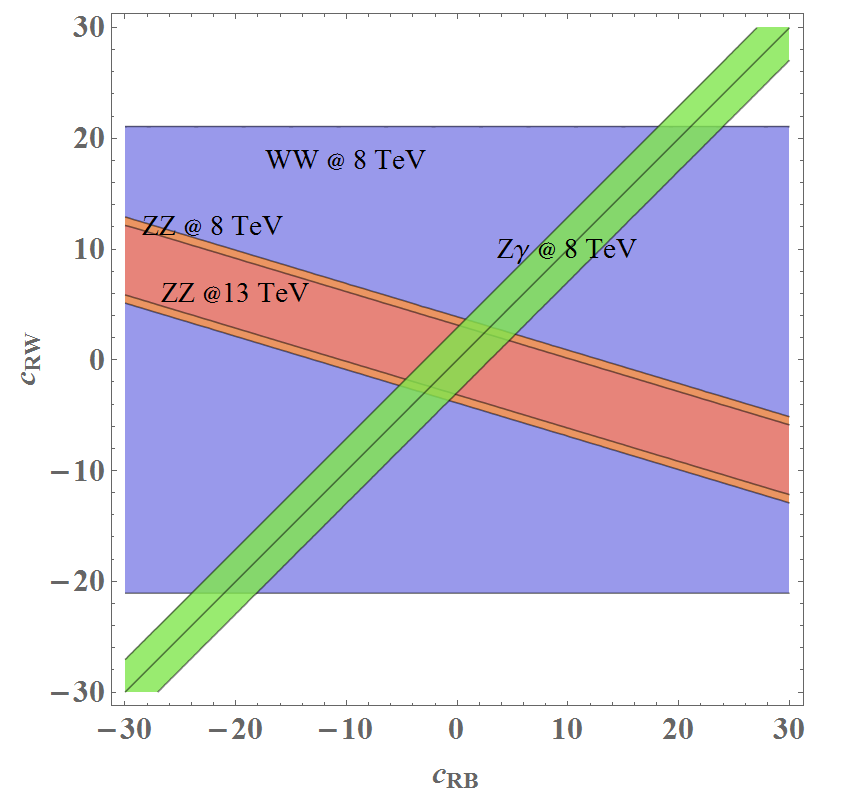}\\
\caption{The allowed parameter spaces for $c_{\namesix B}$ and $c_{\namesix W}$ from $WW$, $ZZ$ and $Z\gamma$ experiment data at $8$ TeV and $13$ TeV LHC.}\label{f_crb}
\end{figure}

According to Eq.~(\ref{eq:lang}), $\namesix$ has many decay channels, and the width of $\namesix$ is
\be
\Gamma(\namesix)=\Gamma(\namesix\to gg)+\Gamma(\namesix\to \nameS\nameS)+\Gamma(\namesix\to \gamma\gamma)+\Gamma(\namesix\to Z\gamma)+\Gamma(\namesix\to ZZ)+\Gamma(\namesix\to WW),
\ee
which depends on the couplings $c_{\namesix W}$, $c_{\namesix B}$, $\shortsixss$ and $\shortsix$.
These parameters will be constrained by the $WW$, $ZZ$ and $Z\gamma$ experiment data at 8 TeV and 13 TeV at the LHC.
When fixing $c_{Rgg}=0.2$ and setting $\Gamma(R)=2~$ GeV, (it is one of the allowed parameters in benchmark 1), the allowed parameter spaces for $c_{\namesix W}$  and $c_{\namesix B}$ from the $WW$, $ZZ$ and $Z\gamma$ production at 8 TeV and 13 TeV LHC~\cite{Aad:2016wpd, ATLAS-CONF-2013-020, Aad:2015zqe, Aad:2016sau} are shown in Fig~\ref{f_crb}. So the constraints for these two coupling are very loose, and there are also very large parameter spaces for other benchmark sets.
In order to simplify the discussion, we choose $c_{\namesix W}=0$. As a result, $\Gamma(\namesix\rightarrow ZZ)\sim0.09~\Gamma(\namesix\rightarrow \gamma\gamma)$ and $\Gamma(\namesix\rightarrow \gamma Z) \sim 0.6~\Gamma(\namesix\rightarrow \gamma\gamma)$, so the width of $\namesix$ can be simplified as
\begin{eqnarray}
\Gamma(\namesix)&=&\Gamma(\namesix\to gg)+\Gamma(\namesix\to \nameS\nameS)+1.69~\Gamma(\namesix\to \gamma\gamma).
\end{eqnarray}
The 1.6 TeV diphoton cross section is
\begin{equation}
\sigma(gg\to\namesix\to\gamma\gamma)=\frac{f_{gg}\Gamma(R)\shortsix}{M_{\namesix}  S_{c.m.}} ~\Gamma(\namesix\to \gamma\gamma)\leq 1\fb.
\end{equation}
We list the  corresponding parameters  to generate $1$ fb cross section on five benchmark sets in Table~\ref{t_aa}. It is obviously that the branch ratio of $\namesix\to\gamma\gamma(ZZ/WW/Z\gamma)$ is small in the parameter spaces where we considered. Because we only need the total decay width of $\namesix$ in the following analysis, as a result, we can safely set $\shortsixaa=0$  in the following discussing.

\begin{table}[h]
\begin{center}
\begin{tabular}{|c|c|c|c|c|c|}
  \hline
  benchmark sets & 1 & 2 & 3 & 4 & 5 \\
  \hline
  $c_{\namesix B}$ & 0.15 & 0.15 & 0.36 & 0.36 & 0.32 \\
  \hline
  $\shortsixaa$ & 0.04 & 0.01 & 0.08 & 0.03 & 0.04 \\
  \hline
  $\Gamma_{\namesix} (\text{GeV})$ & 1.79 & 5.42 & 5.42 & 16.38 & 8.16 \\
  \hline
\end{tabular}
\end{center}
  \caption{The parameter spaces of $c_{\namesix B}$, $\shortsixaa$ and total decay width of $\namesix$  on each benchmark point. The results need to satisfy the $10$ fb diphoton excess.}\label{t_aa}
\end{table}

\begin{figure}[h]
\begin{minipage}[t]{0.33\linewidth}
\centering
 \includegraphics[width=1.0\linewidth]{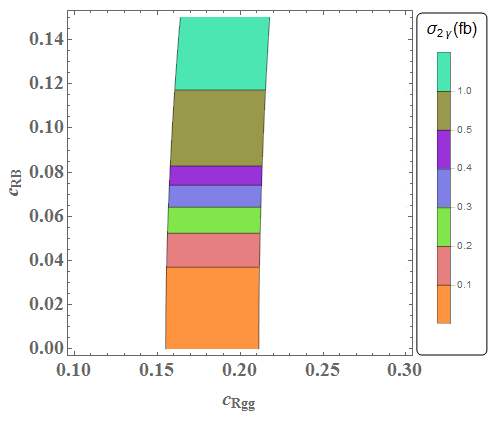}\\
   \text{(a)}
\end{minipage}
\hfill
\begin{minipage}[t]{0.33\linewidth}
\centering
 \includegraphics[width=1.0\linewidth]{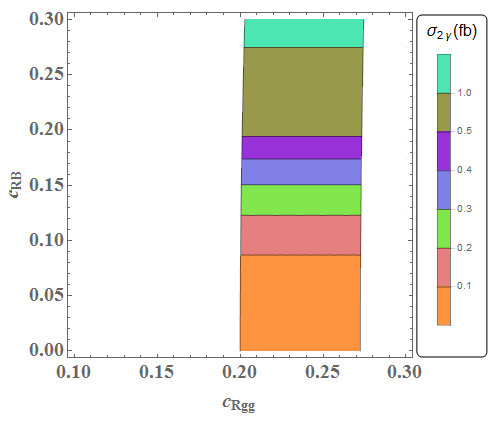}\\
   \text{(b)}
\end{minipage}
\hfill
\begin{minipage}[t]{0.33\linewidth}
\centering
 \includegraphics[width=1.0\linewidth]{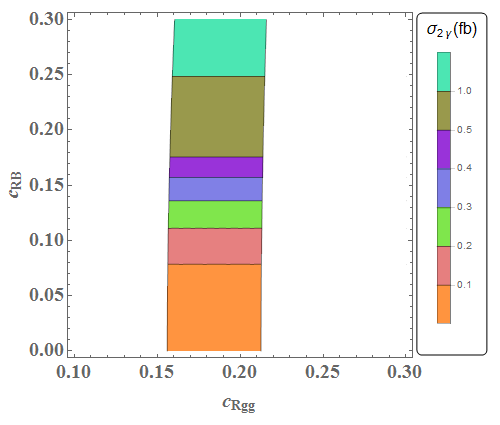}\\
   \text{(c)}
\end{minipage}
\caption{The predicted $gg\rightarrow \namesix \rightarrow 2\gamma$ cross section, where we use benchmark 1 in Figure (a), benchmark 4 in Figure (b) and  benchmark 5 in Figure (c). The constraints mainly come from that we require the cross section $gg\rightarrow \namesix \rightarrow \nameS\nameS\rightarrow 2\gamma 2A$ is about $10\pm 3$ fb. Other constraints on the couplings discussed above are also included.
}\label{f_2gamma}
\end{figure}

In Fig.\ref{f_2gamma}, we present the $gg\rightarrow \namesix \rightarrow \gamma \gamma$ cross sections on $c_{\namesix gg}$-$c_{\namesix B}$ plane, and roughly assume that the cross section for $p p \rightarrow \namesix \rightarrow \gamma\gamma$ is smaller than $1$~fb without any kinematic cuts, which is consistent with current ATLAS diphoton experiment~\cite{ATLAS-CONF-2015-081}. The parameters are also required to generate $10\pm3$~fb diphoton excess at 750~GeV, and we choose benchmark 1, 4  and  5 in Fig.\ref{f_2gamma} (a), (b) and (c). We can see that $\sigma(\namesix\rightarrow \gamma \gamma)$ weakly correlates to $c_{\namesix gg}$, but highly depends on $c_{\namesix B}$ (which also affects $\shortsix$) and $\shortS$.

\begin{figure}[h]
\begin{minipage}[t]{0.43\linewidth}
\centering
 \includegraphics[width=1.0\linewidth]{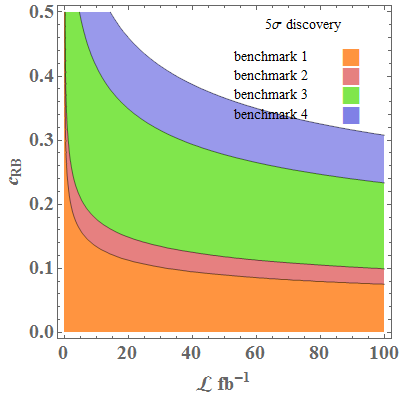}\\
   \text{(a)}
\end{minipage}
\hfill
\begin{minipage}[t]{0.43\linewidth}
\centering
 \includegraphics[width=1.0\linewidth]{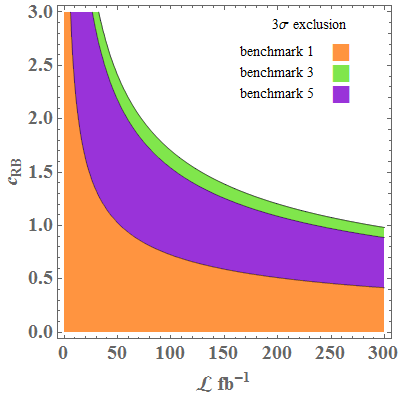}\\
   \text{(b)}
\end{minipage}
  \caption{The 5~$\sigma$ discovery (Fig. (a)) and 3~$\sigma$ exclusion (Fig. (b)) parameter spaces for  $gg\rightarrow \namesix \rightarrow 2\gamma$ in the $\mathcal{L}-c_{\namesix\gamma\gamma}$ plane. Different colors stand for choosing different $\shortsix$, $\shortS$ benchmark points and the corresponding $c_{\namesix gg}$, which satisfy that 750 GeV diphoton cross section is 10 fb in each benchmark points.}\label{f_sig2gamma}
\end{figure}

The simulations of the 1.6 TeV diphoton signal and the corresponding backgrounds are obtained by using the  MadGraph5~\cite{Alwall:2014hca}, with
the cuts in the experiment report~\cite{ATLAS-CONF-2015-081}.
We show the 5 $\sigma$ discovery sensitivities for  $gg\rightarrow \namesix \rightarrow 2\gamma$  with different luminosities in Fig.~\ref{f_sig2gamma} (a). We choose several benchmark points as stated above, and the corresponding $c_{\namesix gg}$ is set to satisfy that the 750 GeV diphoton excess cross section just equals 10 fb in each benchmark point. From the figure, we can see that when $\mathcal{L}\sim 20 ~\text{fb}^{-1}$,  the 1.6 TeV diphoton excess can be observed at the level of 5 $\sigma$ if $c_{\namesix B}$ is larger than 0.5 in all four benchmark points.  In Fig.~\ref{f_sig2gamma} (b), we show the 3 $\sigma$ exclusion limits.  From the Fig.~\ref{f_sig2gamma} (b), we can see that it is hard to exclude $c_{\namesix B}$ to $\mathcal{O}(1)$ even when $\mathcal{L}<100~ \text{fb}^{-1}$ at 13 TeV LHC. So higher luminosity at 14 TeV LHC is needed to exclude this signal.

\subsection{DM relic density}
\begin{figure}
  \centering
  \includegraphics[width=.33\textwidth]{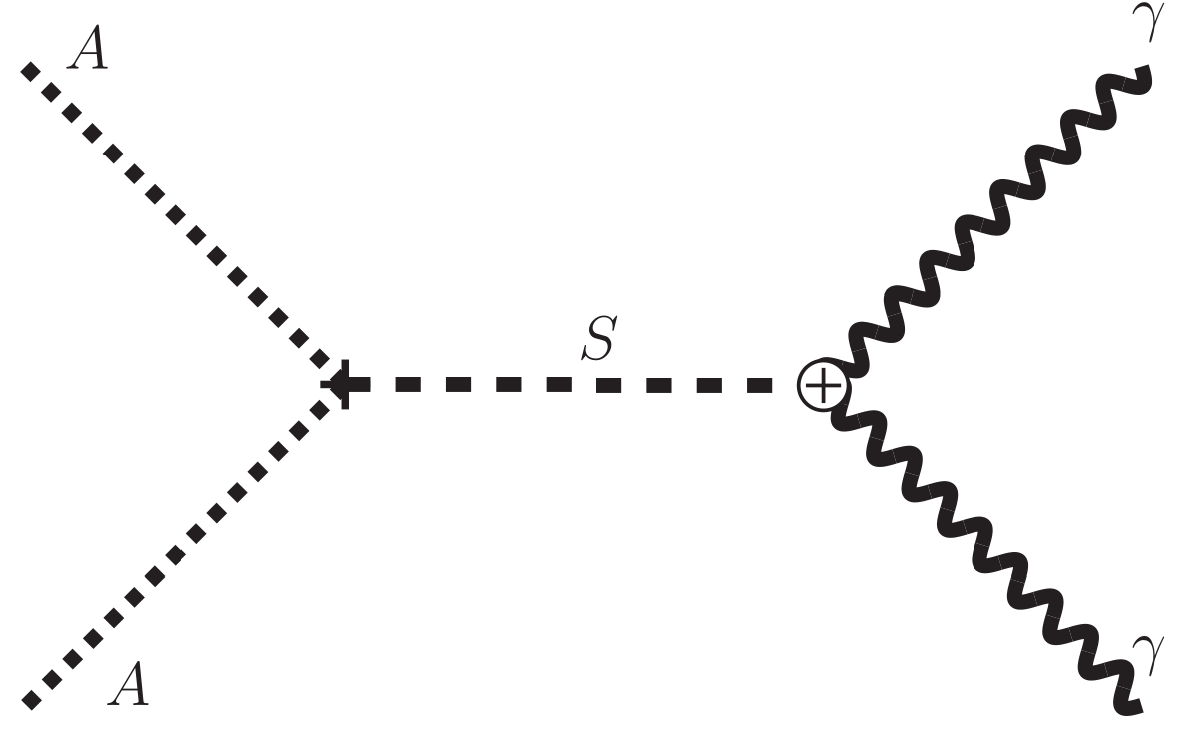}\\
  \caption{The Feynman diagram for dark matter annihilation channel $AA\to\gamma\gamma$.}\label{f_fm_dmaa}
\end{figure}

Since we assume that $\namedm$ is an invisible stable particle, it can be a
natural DM candidate. The Feynman diagram for the dark matter annihilation channel $AA\to\gamma\gamma$ is shown in Fig.~\ref{f_fm_dmaa}.
And we should  study whether the particle $\namedm$ can satisfy the constraint from the observed DM relic density.
Firstly, we obtain the relative velocity ($v$) expanded DM annihilation cross section  $\sigma_{\rm ann} v(\namedm \namedm \to \gamma \gamma) = a+ b v^2 + \mathcal{O}(v^4)$
with
\begin{equation}\label{a_dm}
a=\frac{8 c_{\nameS \namedm\namedm}^2 c_{\nameS \gamma \gamma}^2  M_{\namedm}^2}{\pi((M_{\nameS}^2-4  M_{\namedm}^2 )^2+M_{\nameS}^2 \Gamma_{\nameS}^2)},
\end{equation}
\begin{equation}\label{b_dm}
b=\frac{16 c_{\nameS \namedm\namedm}^2 c_{\nameS \gamma \gamma}^2  M_{\namedm}^4 (M_{\nameS}^2 -4 M_{\namedm}^2)}{\pi((M_{\nameS}^2-4  M_{\namedm}^2)^2+M_{\nameS}^2 \Gamma_{\nameS}^2)^2}.
\end{equation}
Then, from the Boltzmann equation, the corresponding DM relic density can be approximated as~\cite{Jungman:1995df}
\begin{equation}
  \Omega h^2 \simeq \frac{1.04\times 10^9
    x_f   (T_0/2.725~ \rm K)^3 \rm GeV^{-1} }{M_{\rm pl}\sqrt{g_\star (x_f)}(a+ 3
b/x_f)},
\end{equation}
where $x_f \equiv M_{\namedm}/T_f$ with the DM free-out temperatue $T_f$,  $T_0$ is the
cosmic microwave background  temperature at present, and
$g_\star  $ is defined as the effective relativistic
degrees of freedom at $T_f$.

\begin{figure}[h]
\centering
\includegraphics[width=0.45\linewidth]{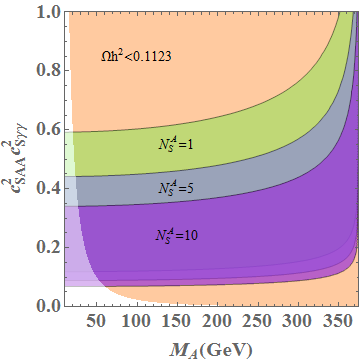}\\
\caption{The constraints for $M_{\namedm}$ and $c_{\nameS\gamma\gamma}^2 c_{\nameS \namedm \namedm}^2$. The pink region stands for the parameter spaces satisfying the observed DM relic density $\Omega h^2 <(0.1186 \pm 0.0020)$~\cite{Ade:2015xua}, 
when the DM is composed of  more than one  species of particles.
The green, blue and purple bands are allowed parameter spaces for $\swidthmin~\text{GeV}<\Gamma(\nameS)<\swidthmax~\text{GeV}$ with $\shortS=1,5$ and $10$, respectively.
So the overlapping regions between the pink and green (blue, purple) ones are the allowed parameter spaces for $\shortS=1$~($5$,~$10$).
}\label{f_dm1}
\end{figure}
In Fig.~\ref{f_dm1}, we show the allowed parameter spaces for the mass and the couplings of $\namedm$. First, the particle $\namedm$ needs to satisfy the current width constraints of $\nameS$, i.e. $\swidthmin\gev<\Gamma(\nameS)<\swidthmax\gev$. We use the green, blue and purple bands denoting the  allowed parameter spaces for the cases of $\shortS=1,5$ and $10$, respectively.
Then, the particle $\namedm$ should contribute to the observed DM relic density. The pink region stands for the parameter spaces satisfy $\Omega h^2 <(0.1186 \pm 0.0020)$~\cite{Ade:2015xua}, if the DM is composed of  more than one  species of particles. And the overlapping regions between the pink and green (blue, purple) ones are the allowed parameter spaces for $\shortS=1$~($5$,~$10$).
Thus, the scalar particle $\namedm$ can be a natural DM candidates with large parameter spaces
in this scenario.
In fact, the particle $\namedm$ can also be a pseudoscalar, a Dirac fermion or a Majorana fermion and so on.
We leave the detailed study on these cases in our future work.
\subsection{the four-photon predictions}
\begin{figure}
  \centering
  \includegraphics[width=.33\textwidth]{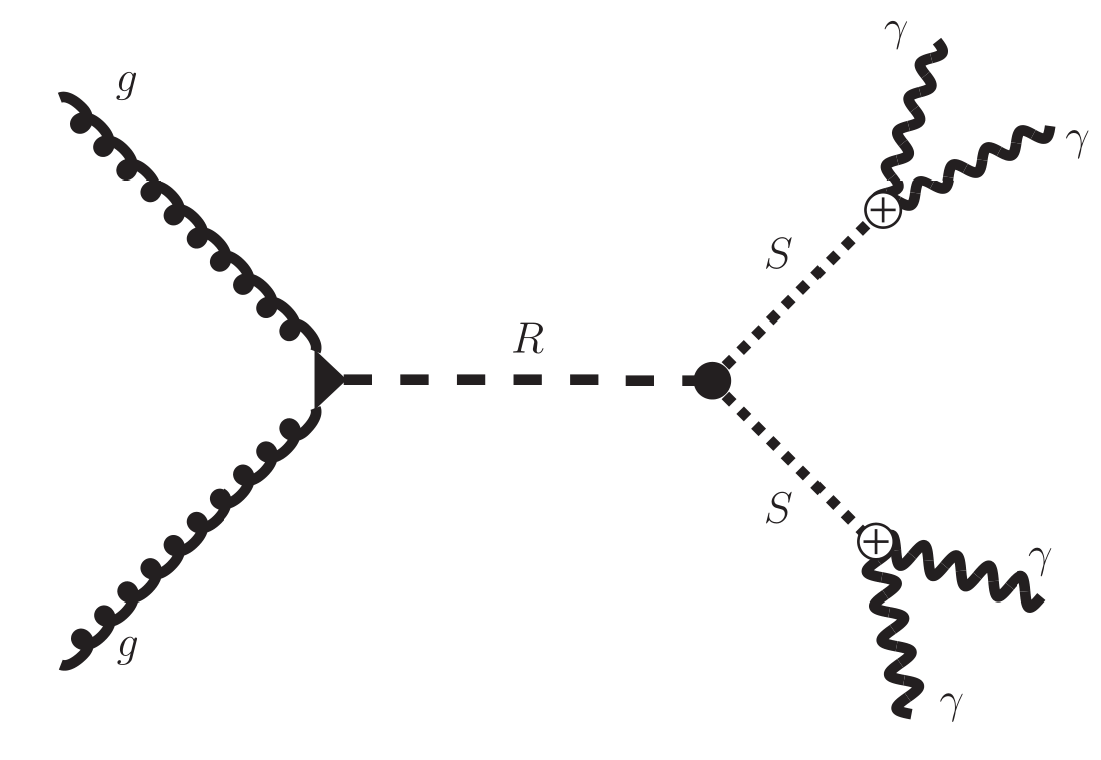}\\
  \caption{The Feynman diagram for the production of the four-photon signal at 1.6 TeV.}\label{4gamma}
\end{figure}
\begin{figure}[h]
\begin{minipage}[t]{0.33\linewidth}
\centering
 \includegraphics[width=1.0\linewidth]{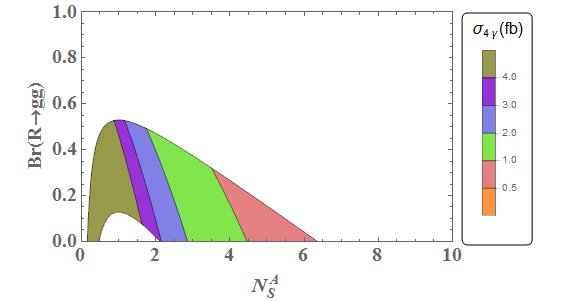}\\
 \text{(a)}
\end{minipage}
\hfill
\begin{minipage}[t]{0.33\linewidth}
\centering
 \includegraphics[width=1.0\linewidth]{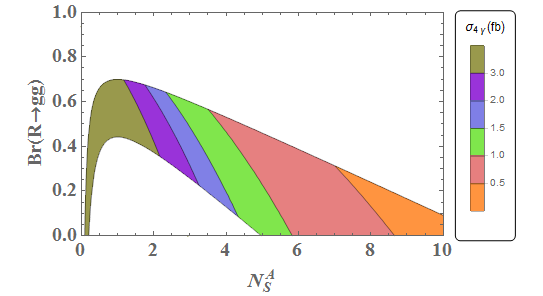}\\
 \text{(b)}
\end{minipage}
\hfill
\begin{minipage}[t]{0.33\linewidth}
\centering
 \includegraphics[width=1.0\linewidth]{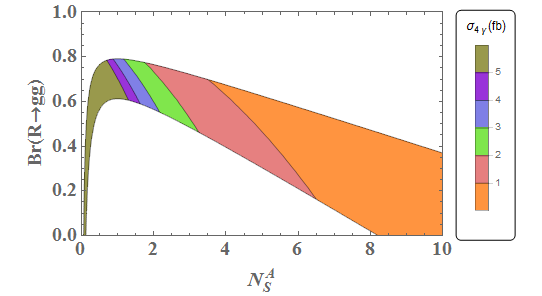}\\
 \text{(c)}
\end{minipage}
 \caption{The predicted $gg\rightarrow 4\gamma$ cross section in $\shortS-\shortsix$ plane with  $c_{\namesix gg}=0.16, 0.2$ and $0.24$, respectively. The cross section for $gg\rightarrow \namesix \rightarrow \nameS\nameS\rightarrow 2\gamma 2\namedm$ is required to be between $7$-$13$ fb. Other constraints from the width of $\nameS$ and $\namesix$, and dijet resonances searching are also included.}\label{f_4gamma}
\end{figure}

In this scenario, the new scalar $\namesix$  also decays to four photons (The corresponding Feynman diagram is in shown in Fig.~\ref{4gamma}.), with the small cross
section
\begin{equation}\label{foura}
\sigma_{4\gamma} \approx \frac{\sigma_{\text{excess}}}{2 \shortS}\approx \frac{5}{\shortS}~\rm fb.
\end{equation}
So the $\sigma_{4\gamma}\sim \mathcal{O}(1)~\text{fb}$.
The  four-photon background in the SM is $\mathcal{O}(10^{-6})~\text{fb}$  in the range of $1550~\text{GeV}<M_{4\gamma}<1650~\text{GeV}$. Thus,
at high luminosity LHC, this new signal is significant, if it is observed.
The numerical results for $c_{\namesix gg}=0.16,~ 0.20$ and 0.24 are shown in Fig.~\ref{f_4gamma}, wherein the cross section for $gg\rightarrow \namesix \rightarrow \nameS\nameS\rightarrow 2\gamma 2\namedm$
is required to be between $7$~fb and $13$~fb, and the constraints from the width of the particle $\nameS$ and $\namesix$,
and dijet resonances searching are also included.

\section{parameter spaces in Case II and Case III}\label{sec:CaseII}
\begin{figure}[h]
\begin{minipage}[t]{0.33\linewidth}
\centering
 \includegraphics[width=1.0\linewidth]{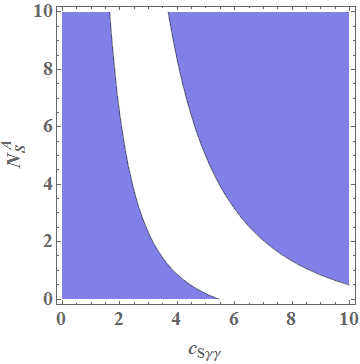}\\
   \text{(a)}
\end{minipage}
\hfill
\begin{minipage}[t]{0.33\linewidth}
\centering
 \includegraphics[width=1.0\linewidth]{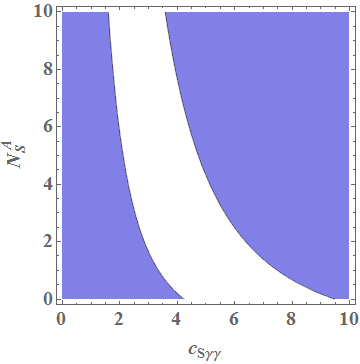}\\
   \text{(b)}
\end{minipage}
\hfill
\begin{minipage}[t]{0.33\linewidth}
\centering
 \includegraphics[width=1.0\linewidth]{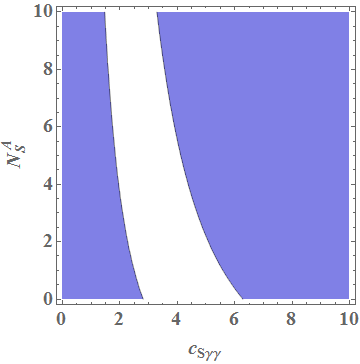}\\
   \text{(c)}
\end{minipage}
 \caption{The constraint of $c_{\nameS\gamma\gamma}$ and $\shortS$ from the width of $\nameS$,
where we assume that the $\nameS$ width is  between $\swidthmin\gev\sim \swidthmax\gev$. The white region corresponds to allowed parameter spaces. (a):~$\nameS$ only couple with $\gamma\gamma$. (b):~Including the contribution from the interaction of $\nameS ZZ$ and $\nameS Z\gamma$ with $c_{\nameS W}=0$. (c):~Including the contribution from the interaction of $\nameS ZZ$ and $\nameS WW$ with $c_{\nameS W}=c_{\nameS B}$.}\label{f_swidth}
\end{figure}
In this section, we consider the Case II and Case III in details.
The most direct variations in these three cases are the allowed parameter spaces from the width of $\nameS$. In Fig.\ref{f_swidth}, we show the constraints of $c_{\nameS\gamma\gamma}$ and $\shortS$ from the width of $\nameS$,
where we assume that $\swidthmin~\rm GeV< \Gamma_{\nameS}< \swidthmax$~GeV. The white regions correspond to the allowed parameter spaces.
The left figure represents Case I (ignore $\nameS WW$, $\nameS ZZ$ or $\nameS Z\gamma$), the middle figure represents Case II (including the contribution from the interaction of $\nameS ZZ$ and $\nameS Z\gamma$ with $c_{\nameS W}=0$.), and the right figure represents Case III (including the contribution from the interaction of $\nameS ZZ$ and $\nameS WW$ with $c_{\nameS W}=c_{\nameS B}$).
 We can see that the three figures are very similar, except that the $c_{\nameS\gamma\gamma}$ varies obviously in Case III when $\shortS$ is small. As a result,
in general, the discussion of the parameter spaces are also applicable in Case II and Case III. But some variation will affect possible parameter spaces for diphoton excess and
other signals. For simplicity, we also ignore the interaction of $\namesix \gamma\gamma$, $\namesix ZZ$, $\namesix Z\gamma$ and  $\namesix WW$ in this section.

\subsection{diphoton excess at 750 GeV}
\begin{figure}[h]
\begin{minipage}[t]{0.43\linewidth}
\centering
 \includegraphics[width=1.0\linewidth]{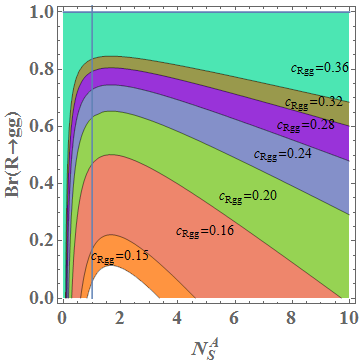}\\
  \text{(a)}
\end{minipage}
\hfill
\begin{minipage}[t]{0.43\linewidth}
\centering
 \includegraphics[width=1.0\linewidth]{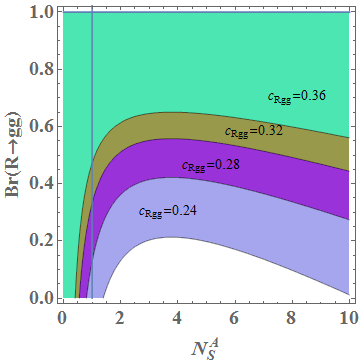}\\
  \text{(b)}
\end{minipage}
\caption{The constraints for $\shortS$ and $\shortsix$ when fixing $c_{\namesix gg}$. The meaning of the colored region is similar with Fig.~\ref{f_nganr}, but we include the contribution from the interaction of $\nameS ZZ$ and $\nameS Z\gamma$  of Case II in Figure (a), and the interaction of $\nameS ZZ$ and $\nameS WW$  of Case III in Figure (b), respectively.
}\label{f_nganr2}
\end{figure}

In Fig.~\ref{f_nganr2}, we plot the constraints for $\shortS$ and $\shortsix$ for Case II and Case III when fixing $c_{\namesix gg}$.  Comparing to the Fig.~\ref{f_nganr}, the allowed regions in Fig.~\ref{f_nganr2} are smaller, because the branch ratio of $\nameS\rightarrow \gamma\gamma$ becomes smaller for the same $\shortS$ in Case II and Case III. So in order to produce the same cross section for diphoton excess, it needs larger cross section of $\namesix \to gg$  production, which leads to a larger $\shortsix$ and $c_{\namesix gg}$. For Case II, extra decay channels lead to $\Gamma(\nameS \rightarrow ZZ)+\Gamma(\nameS \rightarrow Z\gamma) \sim 0.7~\Gamma(\nameS\rightarrow \gamma\gamma)$, so the lower bound for $c_{\namesix gg}^{min}$ is 0.15 in Fig.~\ref{f_nganr2}, instead of $c_{\namesix gg}^{min}\sim 0.12$ in the Fig.~\ref{f_nganr}.  For Case III, the extra decay channels lead to $\Gamma(\nameS \rightarrow ZZ)+\Gamma(\nameS \rightarrow WW) \sim 3~\Gamma(\nameS\rightarrow \gamma\gamma)$, as a result, the remnant parameter spaces are highly suppressed.

\subsection{diboson prediction}
\begin{figure}[h!t]
  \centering
  \includegraphics[width=.33\textwidth]{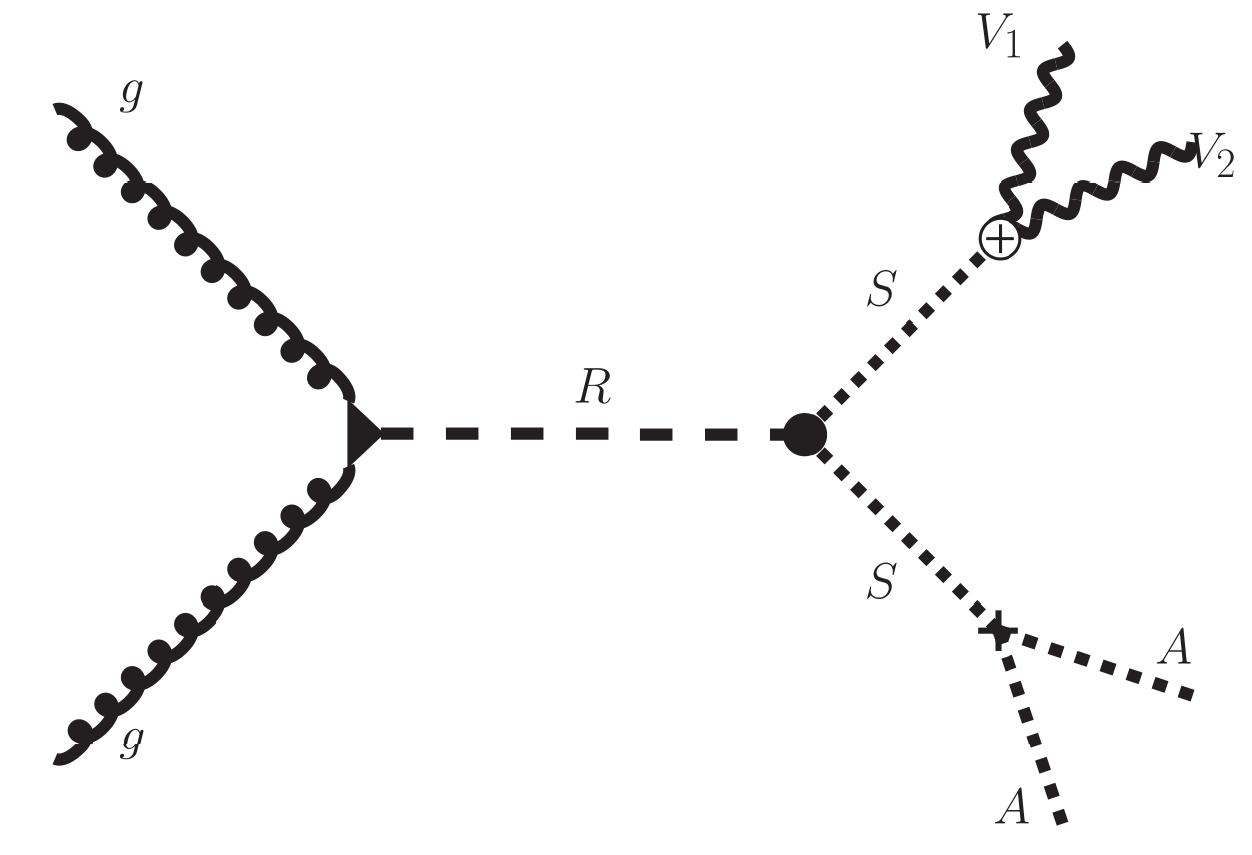}\\
  \caption{The Feynman diagram for di-boson production, where $V_{1}V_{2}=ZZ/Z\gamma/\gamma\gamma$ in Case II and $V_{1}V_{2}=WW/ZZ/\gamma\gamma$ in Case III.}\label{f_fm_2boson}
\end{figure}

In these two cases, similar with diphoton excess, there will also be di-boson signal.
Notice that, for Case II, the cross section of $p p \rightarrow Z\gamma \namedm \namedm$ is close to diphoton channel, where $\Gamma(\nameS\rightarrow Z\gamma) \sim 0.6~\Gamma(\nameS\rightarrow \gamma\gamma)$. And for Case III, the $p p \rightarrow W W \namedm \namedm$ and $p p \rightarrow Z Z \namedm \namedm$ channels become equally significant with the diphoton channel since $\Gamma(\nameS\rightarrow WW) \sim 2~\Gamma(\nameS\rightarrow ZZ) \sim 2 \Gamma(\nameS\rightarrow \gamma\gamma)$. As a result, through the similar procedure as we discussed in the diphoton excess, there will be $Z\gamma$, $WW$ and $ZZ$ signals without large missing energy at a mass of $750$ GeV when choosing different $c_{\nameS B}$ and $c_{\nameS W}$. The Feynman diagram is shown in Fig.\ref{f_fm_2boson}, where $V_{1}V_{2}=ZZ/Z\gamma/\gamma\gamma$ in Case II and $V_{1}V_{2}=WW/ZZ/\gamma\gamma$ in Case III.

Since the $W$ and $Z$ bosons come from the heavy resonance decay, they are highly boosted and form a fat jet. We can identify them with jet substructure. So the collider signal would be a fat $Z$ jet and photon for Case II and fat $W/Z$ jets for Case III. The main SM background would be $\gamma~+$ jets, $Z+\gamma$ and $W+\gamma$ for $Z\gamma \namedm \namedm$ final states and $W/Z~+$ jets, $WW/ZZ/WZ$ pairs, $t\bar{t}$ pairs for $WW/ZZ + \namedm\namedm$ final states. We generate our scenario with Feynrules~\cite{Alloul:2013bka}, and simulate the signals $p p \rightarrow Z\gamma \namedm\namedm$,
$p p \rightarrow Z Z \namedm \namedm$ and $p p \rightarrow W W \namedm \namedm$  and corresponding backgrounds with MadGraph5~\cite{Alwall:2014hca} + Pythia6~\cite{Sjostrand:2006za}. The jet substructure is analyzed with mass-drop technique~\cite{Butterworth:2008iy} and the V-jets ($V = W/Z$) are reconstructed using Cambridge/Aachen algorithm with Fastjet~\cite{Cacciari:2011ma}.

In order to discard the irrelevant backgrounds, we consider the following kinetic cuts.  Firstly, we require $p_{T,J}^{hardest}> 250$ GeV for the hardest jet and $H_T > 500~\text{GeV}$ for the scalar sum of the transverse momentum of final visible states.  The distance parameter R is set as $R=1.2$ to cluster the fat $V$ jets.  The invariant mass of reconstructed V jets satisfy $|M_J-M_V|\leq 13$ GeV. When searching $WW/ZZ + \namedm \namedm$ signal,
since the signal events are highly boosted, we also require $p_{T,J_V}\geq 250$ GeV, $|\eta_{J_V}|\leq 3$. The invariant mass of two reconstructed V-jets should satisfy $|M_{J_{V1}J_{V2}}-M_{\nameS}|\leq 50$ GeV. When searching the $Z\gamma \namedm \namedm$ signal, we require a tagged hard photon, which satisfies $p_{T,\gamma}\geq 250$ GeV, $|\eta_{\gamma}|\leq 1.4$, and the invariant mass of $\gamma$ and the $Z$-tagged jet should satisfy $|M_{\gamma J_Z}-M_{\nameS}|\leq 25$ GeV.
\begin{figure}[h!t]
\begin{minipage}[t]{0.33\linewidth}
\centering
 \includegraphics[width=1.0\linewidth]{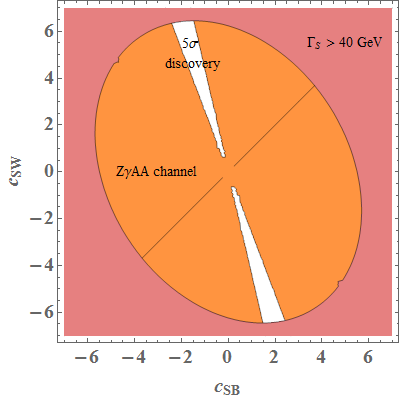}\\
  \text{(a)}
\end{minipage}
\hfill
\begin{minipage}[t]{0.33\linewidth}
\centering
 \includegraphics[width=1.0\linewidth]{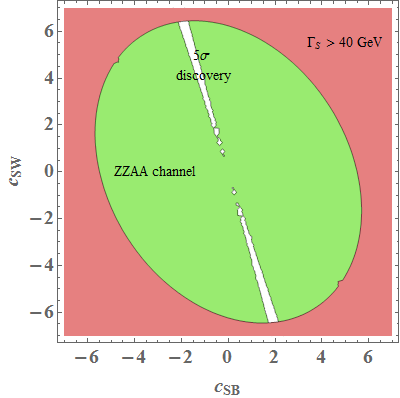}\\
  \text{(b)}
\end{minipage}
\hfill
\begin{minipage}[t]{0.33\linewidth}
\centering
 \includegraphics[width=1.0\linewidth]{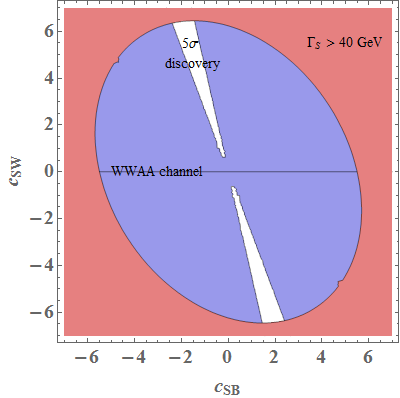}\\
  \text{(c)}
\end{minipage}
 \caption{The 5~$\sigma$ discovery potential for $g g\rightarrow \namesix \rightarrow \nameS\nameS \rightarrow Z\gamma \namedm \namedm/Z Z \namedm \namedm/W W \namedm \namedm$ in the $c_{\nameS B}-c_{\nameS W}$ plane with 100 $fb^{-1}$ at the 13 TeV LHC. The parameter spaces are consistent with the diphoton excess cross section, i.e.  $\sigma_{\gamma\gamma \namedm\namedm}=10~\text{fb}$. The pink regions stand for that the width of $\nameS$ is larger than $40$ GeV. The white regions in three figures are where the signal can be observed over the level of 5~$\sigma$.}\label{f_dis_WWZZ}
\end{figure}
We show the discovery potentials of the above signals with an integrated luminosity of 100 $\text{fb}^{-1}$ at the 13 TeV LHC in Fig.~\ref{f_dis_WWZZ}. The parameter spaces are required to be consistent with the diphoton excess cross section, i.e.  $\sigma_{\gamma\gamma\namedm \namedm}=10~\text{fb}$. In the figures, the pink regions stand for that the width of $\nameS$ is larger than $40$ GeV and the white regions  mean that the signal can be observed over the level of 5~$\sigma$. The signals can be observed when  $c_{\nameS B}$ and   $c_{\nameS W}$ are with the opposite signs as well as $|c_{\nameS W}/c_{\nameS B}|$ is large. The three figures look very similar and most of the parameter spaces are excluded,  because the backgrounds are too large for the all three signals. So these signals can only  be observed if the parameters are more like Case II at the 13 TeV LHC. But these signals in larger parameter spaces may be observed at the  high luminosity 14 TeV LHC.

\subsection{DM relic density}
\begin{figure}[h!t]
  \centering
  \includegraphics[width=.33\textwidth]{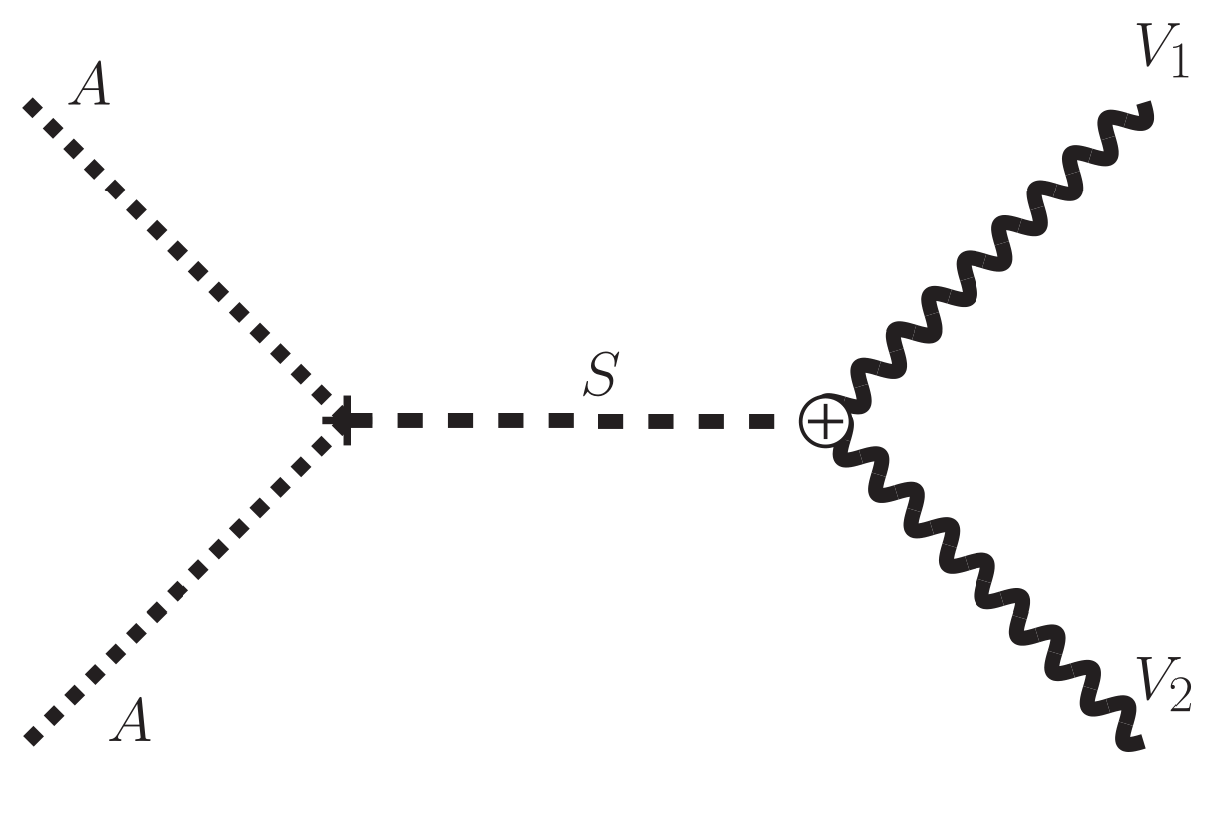}\\
  \caption{The Feynman diagram for dark matter annihilation channel $AA\to V_{1}V_{2}$, where $V_{1}V_{2}=\gamma\gamma,~Z\gamma,~ZZ$ and $WW$.
  In case II, $V_{1}V_{2}=\gamma\gamma,~Z\gamma~\rm and~ZZ$;
  In case III,  $V_{1}V_{2}=\gamma\gamma,~ZZ$ and $WW$.  }\label{f_fm_dmvv}
\end{figure}

\begin{figure}[h]
\begin{minipage}[t]{0.43\linewidth}
\centering
 \includegraphics[width=1.0\linewidth]{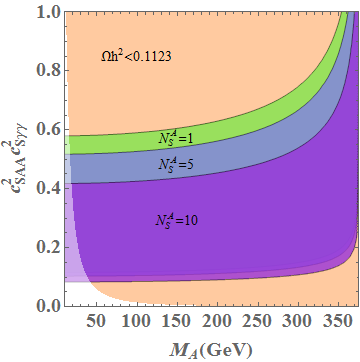}\\
  \text{(a)}
\end{minipage}
\hfill
\begin{minipage}[t]{0.43\linewidth}
\centering
 \includegraphics[width=1.0\linewidth]{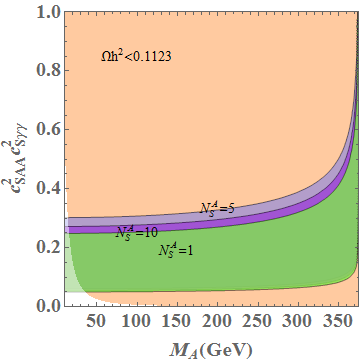}\\
  \text{(b)}
\end{minipage}
\caption{The constraints for $M_{\namedm}$ and $c_{\nameS\gamma\gamma}^2 c_{\nameS \namedm \namedm}^2$ including the constraints of the DM relic density.
The meaning of the colored region is similar with Fig.~\ref{f_dm1}, but we include the contribution from the interaction of $\nameS ZZ$ and $\nameS Z\gamma$  of Case II in Figure (a), and the interaction of $\nameS ZZ$ and $\nameS WW$  of Case III in Figure (b), respectively.
}\label{f_dm2}
\end{figure}

In Case II and III, the  Feynman diagram for the dark matter annihilation $AA\to V_{1}V_{2}$ is shown in Fig.~\ref{f_fm_dmvv},
where
$V_{1}V_{2}=\gamma\gamma,~Z\gamma~\rm and~ZZ$ in case II;
$V_{1}V_{2}=\gamma\gamma,~ZZ$ and $WW$ in case III.
Here, we only show the numerical results, and the couplings $c_{SZ\gamma}$, $c_{SWW}$ and
$c_{SZZ}$ are written in terms of $c_{S\gamma\gamma}$.
Namely, $c_{SZ\gamma}\approx 0.3 ~c_{S\gamma\gamma}$, $c_{SZ\gamma}\approx~ 0.78 c_{S\gamma\gamma}$ in Case II;
and  $c_{SZZ}\approx c_{S\gamma\gamma}$, $c_{SWW}\approx 1.41~c_{S\gamma\gamma}$ in Case III.
Figure~\ref{f_dm2} shows the constraints for DM mass $M_{\namedm}$ and  the couplings $c_{\nameS\gamma\gamma}^2 c_{\nameS \namedm \namedm}^2$ after including the constraints from the observed DM relic density.
The meaning of the colored region is similar with Fig.~\ref{f_dm1}, but we include the contribution from the interaction of $\nameS ZZ$ and $\nameS Z\gamma$  of Case II in Figure (a), and the interaction of $\nameS ZZ$ and $\nameS WW$  of Case III in Figure (b), respectively.
Comparing with Fig.~\ref{f_dm1}, the overlapping region become larger when $\shortS=1,5$ and $10$. That is because we define $\shortS=\frac{\Gamma(\nameS \to \namedm \namedm)}{\Gamma(\nameS \to \gamma \gamma)}$, and $\Gamma(\nameS \to Z \gamma/ZZ/WW)$ also contribute to the total decay width for a fixed $\shortS$.
For Case III, due to the large branch ratio of $\Gamma(\nameS\rightarrow WW/ZZ)$,
there are small differences for different $\shortS$.

\subsection{four-boson prediction}
\begin{figure}[h]
\begin{center}
  \text{Case II}
\end{center}
\begin{minipage}[t]{0.33\linewidth}
\centering
 \includegraphics[width=1.0\linewidth]{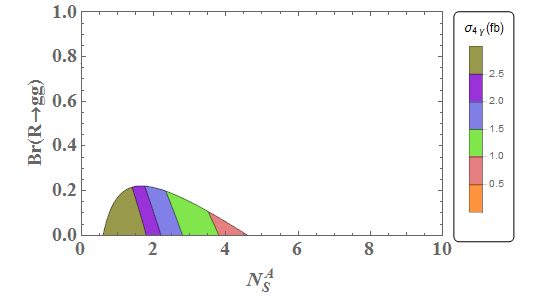}\\
 \text{(a)}
\end{minipage}
\hfill
\begin{minipage}[t]{0.33\linewidth}
\centering
 \includegraphics[width=1.0\linewidth]{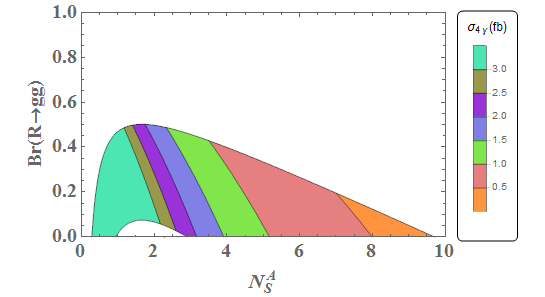}\\
  \text{(b)}
\end{minipage}
\hfill
\begin{minipage}[t]{0.33\linewidth}
\centering
 \includegraphics[width=1.0\linewidth]{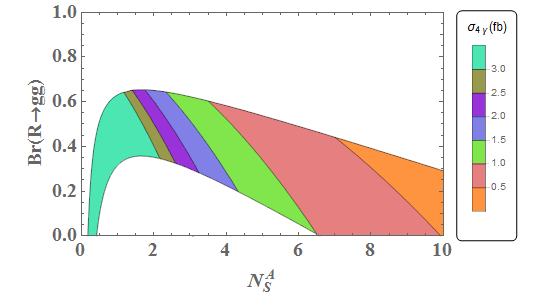}\\
  \text{(c)}
\end{minipage}
\hfill
\begin{center}
  \text{Case III}
\end{center}
\begin{minipage}[t]{0.33\linewidth}
\centering
 \includegraphics[width=1.0\linewidth]{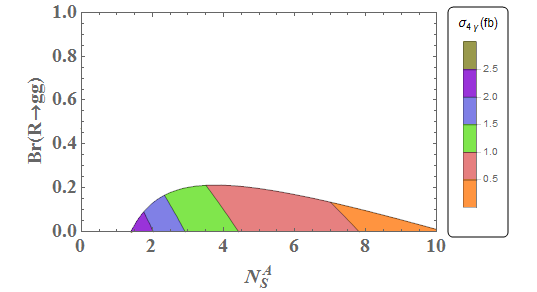}\\
  \text{(d)}
\end{minipage}
\hfill
\begin{minipage}[t]{0.33\linewidth}
\centering
 \includegraphics[width=1.0\linewidth]{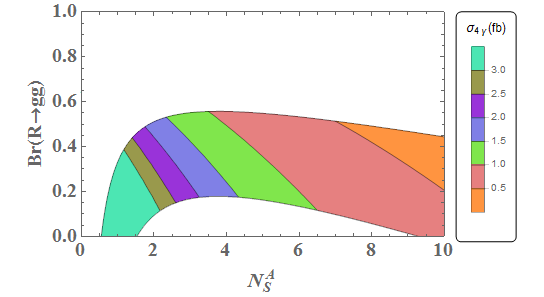}\\
  \text{(e)}
\end{minipage}
\hfill
\begin{minipage}[t]{0.33\linewidth}
\centering
 \includegraphics[width=1.0\linewidth]{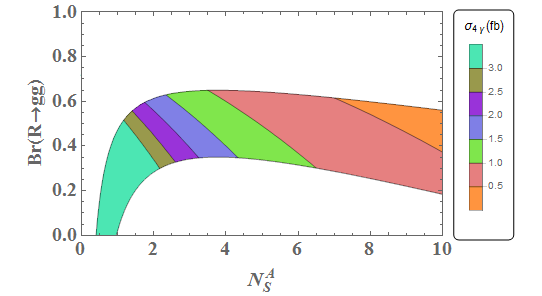}\\
  \text{(f)}
\end{minipage}
 \caption{The predicted $gg\rightarrow 4\gamma$ cross section in $\shortS-\shortsix$ plane with  $c_{\namesix gg}=0.16,~0.20$ and $0.24$ for Case II and  $c_{\namesix gg}=0.24,~0.30$ and $0.34$ for Case III, respectively. The cross section for $gg\rightarrow \namesix \rightarrow \nameS\nameS\rightarrow 2\gamma 2\namedm$ is required to be between $7-13$ fb.  Other constraints from the width of $\nameS$ and $\namesix$, and dijet resonances searching are also included. These figures are similar with Fig.~\ref{f_4gamma}, but we include the contribution from the interaction of $\nameS ZZ$ and $\nameS Z\gamma$ of Case II in the Figures (a),(b), and (c), and
 the interaction of $\nameS ZZ$ and $\nameS WW$  of Case III in the Figures (d),(e), and (f), respectively.}\label{f_4gamma2}
\end{figure}

In Case II and Case III, there are also the four-photon signal.
In Fig.~\ref{f_4gamma2}, we present the predicted $gg\rightarrow 4\gamma$ signals for Case II and Case III, respectively. We choose $c_{\namesix gg}=0.16,~0.20$ and $0.24$ for Case II (Fig.~\ref{f_4gamma2} (a), (b), (c)), and $c_{\namesix gg}=0.24,~0.30$ and $0.34$  for Case III (Fig.~\ref{f_4gamma2} (d), (e), (f)).
The $4\gamma$ cross section turns to be compatible with $750$ GeV diphoton cross section in some allowed parameter spaces, i.e. $\sigma_{4\gamma}\sim \mathcal{O}(1)~\text{fb}$.

However, the best channels in this scenario are four-boson production at a invariant mass of $1.6$ TeV, such as $WWWW$, $WW\gamma\gamma$, $WWZZ$, $WWZ\gamma$, $ZZZZ$, $ZZZ\gamma$, $ZZ\gamma\gamma$ and $Z\gamma\gamma\gamma$ production, which is similar to the four-photon signal discussed above. The cross section of these four-boson signals can reach to $\mathcal{O}(1)~\text{fb}$, while the backgrounds of the four bosons can almost be ignored at 1.6 TeV. However, the cross section of four boson production highly rely on the choice of $c_{\nameS W}$ and $c_{\nameS B}$. As a result, we can search the four-boson signal to constrain the $c_{\nameS W}$ and $c_{\nameS B}$ at the high luminosity LHC.

\section{Discussion and Conclusion}\label{sec:conclusion}
In general, the particle $\namesix$ can also couple with the $Z$ and $W$ bosons, which will
lead to more  abundant signals at 1.6 TeV, such as the signals of  gauge boson pairs.
Since we are only interested in the diphoton and four-boson signals in this paper, we leave the detailed discussion on these
gauge boson signals in our future work.

At the same time, the existence of scalar $\namesix$ can also explain
the diphoton deviation with an invariant mass of  about 1.6 TeV at the ATLAS.
Especially, we predict the slight four-photon and four-boson excess, which can be tested at the high luminosity LHC in the future.
As a by-product, our scenario also naturally provides the  DM candidates.

In August 2016, ATLAS~\cite{ATLAS:2016eeo} released the new data on diphoton resonance\footnote{For the CMS, the roughly 2.9 $\sigma$ excess of the 750 GeV diphoton observed in the 2015 data is reduced to about 0.8 $\sigma$ in the combined analysis of 2015 and 2016 data~\cite{CMS:2016crm}.}.
There exists a 2.4 $\sigma$ deviation over the background-only hypothesis at the 1.6 TeV when the  dataset in 2015 and 2016 are combined.
The 3.9 $\sigma$ excess near 750 GeV in the 2015 data is changed to the 2.3 $\sigma$ deviation near 710 GeV with the width of $70$ GeV  in the
combined analysis of 2015 and 2016 data.
This new result in terms of cross section can be roughly estimated as $\sigma_{\rm excess} \lesssim  3~\text{fb} ~(\rm at~13~TeV ~ATLAS)$, which is about one third of the  ATLAS results in 2015.
However, all the analysis procedure with new data will
be the same as above except that the cross section of diphoton excess is changed from 10 fb to 3 fb.  And, the allowed  parameter spaces and the predicted signals strengths will be suppressed less than an order of magnitude.
Thus, even including the new data, there still leaves enough parameter spaces for this scenario.


In this work, we use the 750 GeV excess as trial data to study the  heavy diphoton excess and dark matter by the cascade decay scenario.
Although it is disfavored by the new data, this cascade scenario may be used to study other possible  heavy resonance, such as dijet resonance and diboson resonance
at LHC with high luminosity.
Namely, the kinematic of decay products of a possible  resonance can be similar with  a heavier particle decaying to that resonance and the
DM, if the mass of the heavier particle is about twice of the resonance mass.  So the signal from the former can be regarded as the signal from the latter at colliders.
This scenario could help us to study the DM and other new physics
beyond the SM at future lepton and hadron colliders.

\begin{acknowledgements}
The authors of this paper gratefully acknowledge  valuable comments from Hao Zhang and Peng-Fei Yin.
Chong Sheng Li and Ze Long Liu are supported by the National Natural Science Foundation of China, under Grants
No.11375013 and No. 11135003.
Fa Peng Huang is supported by the NSFC under grants Nos.11121092, 11033005, 11375220, the CAS pilotB program
and the China Postdoctoral Science Foundation under Grant No. 2016M590133.
\end{acknowledgements}

\bibliography{WY_HFPb}

\begin{thebibliography}{66}
\expandafter\ifx\csname natexlab\endcsname\relax\def\natexlab#1{#1}\fi
\expandafter\ifx\csname bibnamefont\endcsname\relax
  \def\bibnamefont#1{#1}\fi
\expandafter\ifx\csname bibfnamefont\endcsname\relax
  \def\bibfnamefont#1{#1}\fi
\expandafter\ifx\csname citenamefont\endcsname\relax
  \def\citenamefont#1{#1}\fi
\expandafter\ifx\csname url\endcsname\relax
  \def\url#1{\texttt{#1}}\fi
\expandafter\ifx\csname urlprefix\endcsname\relax\def\urlprefix{URL }\fi
\providecommand{\bibinfo}[2]{#2}
\providecommand{\eprint}[2][]{\url{#2}}

\bibitem[{ATL(2015)}]{ATLAS-CONF-2015-081}
\bibinfo{type}{Tech. Rep.} \bibinfo{number}{ATLAS-CONF-2015-081},
  \bibinfo{institution}{The ATLAS Collaboration} (\bibinfo{year}{2015}),
  \urlprefix\url{http://cds.cern.ch/record/2114853}.

\bibitem[{ATL(2016)}]{ATLAS-CONF-2016-018}
\bibinfo{type}{Tech. Rep.} \bibinfo{number}{ATLAS-CONF-2016-018},
  \bibinfo{institution}{The ATLAS Collaboration}, \bibinfo{address}{Geneva}
  (\bibinfo{year}{2016}), \urlprefix\url{http://cds.cern.ch/record/2141568}.

\bibitem[{CMS(2015{\natexlab{a}})}]{CMS-PAS-EXO-15-004}
\bibinfo{type}{Tech. Rep.} \bibinfo{number}{CMS-PAS-EXO-15-004},
  \bibinfo{institution}{The CMS Collaboration}
  (\bibinfo{year}{2015}{\natexlab{a}}),
  \urlprefix\url{http://cds.cern.ch/record/2114808}.

\bibitem[{CMS(2016)}]{CMS-PAS-EXO-16-018}
\bibinfo{type}{Tech. Rep.} \bibinfo{number}{CMS-PAS-EXO-16-018},
  \bibinfo{institution}{The CMS Collaboration} (\bibinfo{year}{2016}),
  \urlprefix\url{http://cds.cern.ch/record/2139899}.

\bibitem[{\citenamefont{Harigaya and Nomura}(2016)}]{Harigaya:2015ezk}
\bibinfo{author}{\bibfnamefont{K.}~\bibnamefont{Harigaya}} \bibnamefont{and}
  \bibinfo{author}{\bibfnamefont{Y.}~\bibnamefont{Nomura}},
  \bibinfo{journal}{Phys. Lett.} \textbf{\bibinfo{volume}{B754}},
  \bibinfo{pages}{151} (\bibinfo{year}{2016}), \eprint{1512.04850}.

\bibitem[{\citenamefont{Mambrini et~al.}(2016)\citenamefont{Mambrini, Arcadi,
  and Djouadi}}]{Mambrini:2015wyu}
\bibinfo{author}{\bibfnamefont{Y.}~\bibnamefont{Mambrini}},
  \bibinfo{author}{\bibfnamefont{G.}~\bibnamefont{Arcadi}}, \bibnamefont{and}
  \bibinfo{author}{\bibfnamefont{A.}~\bibnamefont{Djouadi}},
  \bibinfo{journal}{Phys. Lett.} \textbf{\bibinfo{volume}{B755}},
  \bibinfo{pages}{426} (\bibinfo{year}{2016}), \eprint{1512.04913}.

\bibitem[{\citenamefont{Backovic et~al.}(2016)\citenamefont{Backovic, Mariotti,
  and Redigolo}}]{Backovic:2015fnp}
\bibinfo{author}{\bibfnamefont{M.}~\bibnamefont{Backovic}},
  \bibinfo{author}{\bibfnamefont{A.}~\bibnamefont{Mariotti}}, \bibnamefont{and}
  \bibinfo{author}{\bibfnamefont{D.}~\bibnamefont{Redigolo}},
  \bibinfo{journal}{JHEP} \textbf{\bibinfo{volume}{03}}, \bibinfo{pages}{157}
  (\bibinfo{year}{2016}), \eprint{1512.04917}.

\bibitem[{\citenamefont{Angelescu et~al.}(2016)\citenamefont{Angelescu,
  Djouadi, and Moreau}}]{Angelescu:2015uiz}
\bibinfo{author}{\bibfnamefont{A.}~\bibnamefont{Angelescu}},
  \bibinfo{author}{\bibfnamefont{A.}~\bibnamefont{Djouadi}}, \bibnamefont{and}
  \bibinfo{author}{\bibfnamefont{G.}~\bibnamefont{Moreau}},
  \bibinfo{journal}{Phys. Lett.} \textbf{\bibinfo{volume}{B756}},
  \bibinfo{pages}{126} (\bibinfo{year}{2016}), \eprint{1512.04921}.

\bibitem[{\citenamefont{Nakai et~al.}(2016)\citenamefont{Nakai, Sato, and
  Tobioka}}]{Nakai:2015ptz}
\bibinfo{author}{\bibfnamefont{Y.}~\bibnamefont{Nakai}},
  \bibinfo{author}{\bibfnamefont{R.}~\bibnamefont{Sato}}, \bibnamefont{and}
  \bibinfo{author}{\bibfnamefont{K.}~\bibnamefont{Tobioka}},
  \bibinfo{journal}{Phys. Rev. Lett.} \textbf{\bibinfo{volume}{116}},
  \bibinfo{pages}{151802} (\bibinfo{year}{2016}), \eprint{1512.04924}.

\bibitem[{\citenamefont{Knapen et~al.}(2016)\citenamefont{Knapen, Melia,
  Papucci, and Zurek}}]{Knapen:2015dap}
\bibinfo{author}{\bibfnamefont{S.}~\bibnamefont{Knapen}},
  \bibinfo{author}{\bibfnamefont{T.}~\bibnamefont{Melia}},
  \bibinfo{author}{\bibfnamefont{M.}~\bibnamefont{Papucci}}, \bibnamefont{and}
  \bibinfo{author}{\bibfnamefont{K.}~\bibnamefont{Zurek}},
  \bibinfo{journal}{Phys. Rev.} \textbf{\bibinfo{volume}{D93}},
  \bibinfo{pages}{075020} (\bibinfo{year}{2016}), \eprint{1512.04928}.

\bibitem[{\citenamefont{Buttazzo et~al.}(2016)\citenamefont{Buttazzo, Greljo,
  and Marzocca}}]{Buttazzo:2015txu}
\bibinfo{author}{\bibfnamefont{D.}~\bibnamefont{Buttazzo}},
  \bibinfo{author}{\bibfnamefont{A.}~\bibnamefont{Greljo}}, \bibnamefont{and}
  \bibinfo{author}{\bibfnamefont{D.}~\bibnamefont{Marzocca}},
  \bibinfo{journal}{Eur. Phys. J.} \textbf{\bibinfo{volume}{C76}},
  \bibinfo{pages}{116} (\bibinfo{year}{2016}), \eprint{1512.04929}.

\bibitem[{\citenamefont{Pilaftsis}(2016)}]{Pilaftsis:2015ycr}
\bibinfo{author}{\bibfnamefont{A.}~\bibnamefont{Pilaftsis}},
  \bibinfo{journal}{Phys. Rev.} \textbf{\bibinfo{volume}{D93}},
  \bibinfo{pages}{015017} (\bibinfo{year}{2016}), \eprint{1512.04931}.

\bibitem[{\citenamefont{Franceschini et~al.}(2016)\citenamefont{Franceschini,
  Giudice, Kamenik, McCullough, Pomarol, Rattazzi, Redi, Riva, Strumia, and
  Torre}}]{Franceschini:2015kwy}
\bibinfo{author}{\bibfnamefont{R.}~\bibnamefont{Franceschini}},
  \bibinfo{author}{\bibfnamefont{G.~F.} \bibnamefont{Giudice}},
  \bibinfo{author}{\bibfnamefont{J.~F.} \bibnamefont{Kamenik}},
  \bibinfo{author}{\bibfnamefont{M.}~\bibnamefont{McCullough}},
  \bibinfo{author}{\bibfnamefont{A.}~\bibnamefont{Pomarol}},
  \bibinfo{author}{\bibfnamefont{R.}~\bibnamefont{Rattazzi}},
  \bibinfo{author}{\bibfnamefont{M.}~\bibnamefont{Redi}},
  \bibinfo{author}{\bibfnamefont{F.}~\bibnamefont{Riva}},
  \bibinfo{author}{\bibfnamefont{A.}~\bibnamefont{Strumia}}, \bibnamefont{and}
  \bibinfo{author}{\bibfnamefont{R.}~\bibnamefont{Torre}},
  \bibinfo{journal}{JHEP} \textbf{\bibinfo{volume}{03}}, \bibinfo{pages}{144}
  (\bibinfo{year}{2016}), \eprint{1512.04933}.

\bibitem[{\citenamefont{Di~Chiara et~al.}(2016)\citenamefont{Di~Chiara,
  Marzola, and Raidal}}]{DiChiara:2015vdm}
\bibinfo{author}{\bibfnamefont{S.}~\bibnamefont{Di~Chiara}},
  \bibinfo{author}{\bibfnamefont{L.}~\bibnamefont{Marzola}}, \bibnamefont{and}
  \bibinfo{author}{\bibfnamefont{M.}~\bibnamefont{Raidal}},
  \bibinfo{journal}{Phys. Rev.} \textbf{\bibinfo{volume}{D93}},
  \bibinfo{pages}{095018} (\bibinfo{year}{2016}), \eprint{1512.04939}.

\bibitem[{\citenamefont{Fiore et~al.}(2016)\citenamefont{Fiore, Jenkovszky, and
  Schicker}}]{Fiore:2015lnz}
\bibinfo{author}{\bibfnamefont{R.}~\bibnamefont{Fiore}},
  \bibinfo{author}{\bibfnamefont{L.}~\bibnamefont{Jenkovszky}},
  \bibnamefont{and} \bibinfo{author}{\bibfnamefont{R.}~\bibnamefont{Schicker}},
  \bibinfo{journal}{Eur. Phys. J.} \textbf{\bibinfo{volume}{C76}},
  \bibinfo{pages}{38} (\bibinfo{year}{2016}), \eprint{1512.04977}.

\bibitem[{\citenamefont{Higaki et~al.}(2016)\citenamefont{Higaki, Jeong,
  Kitajima, and Takahashi}}]{Higaki:2015jag}
\bibinfo{author}{\bibfnamefont{T.}~\bibnamefont{Higaki}},
  \bibinfo{author}{\bibfnamefont{K.~S.} \bibnamefont{Jeong}},
  \bibinfo{author}{\bibfnamefont{N.}~\bibnamefont{Kitajima}}, \bibnamefont{and}
  \bibinfo{author}{\bibfnamefont{F.}~\bibnamefont{Takahashi}},
  \bibinfo{journal}{Phys. Lett.} \textbf{\bibinfo{volume}{B755}},
  \bibinfo{pages}{13} (\bibinfo{year}{2016}), \eprint{1512.05295}.

\bibitem[{\citenamefont{McDermott et~al.}(2016)\citenamefont{McDermott, Meade,
  and Ramani}}]{McDermott:2015sck}
\bibinfo{author}{\bibfnamefont{S.~D.} \bibnamefont{McDermott}},
  \bibinfo{author}{\bibfnamefont{P.}~\bibnamefont{Meade}}, \bibnamefont{and}
  \bibinfo{author}{\bibfnamefont{H.}~\bibnamefont{Ramani}},
  \bibinfo{journal}{Phys. Lett.} \textbf{\bibinfo{volume}{B755}},
  \bibinfo{pages}{353} (\bibinfo{year}{2016}), \eprint{1512.05326}.

\bibitem[{\citenamefont{Low et~al.}(2016)\citenamefont{Low, Tesi, and
  Wang}}]{Low:2015qep}
\bibinfo{author}{\bibfnamefont{M.}~\bibnamefont{Low}},
  \bibinfo{author}{\bibfnamefont{A.}~\bibnamefont{Tesi}}, \bibnamefont{and}
  \bibinfo{author}{\bibfnamefont{L.-T.} \bibnamefont{Wang}},
  \bibinfo{journal}{JHEP} \textbf{\bibinfo{volume}{03}}, \bibinfo{pages}{108}
  (\bibinfo{year}{2016}), \eprint{1512.05328}.

\bibitem[{\citenamefont{Bellazzini et~al.}(2016)\citenamefont{Bellazzini,
  Franceschini, Sala, and Serra}}]{Bellazzini:2015nxw}
\bibinfo{author}{\bibfnamefont{B.}~\bibnamefont{Bellazzini}},
  \bibinfo{author}{\bibfnamefont{R.}~\bibnamefont{Franceschini}},
  \bibinfo{author}{\bibfnamefont{F.}~\bibnamefont{Sala}}, \bibnamefont{and}
  \bibinfo{author}{\bibfnamefont{J.}~\bibnamefont{Serra}},
  \bibinfo{journal}{JHEP} \textbf{\bibinfo{volume}{04}}, \bibinfo{pages}{072}
  (\bibinfo{year}{2016}), \eprint{1512.05330}.

\bibitem[{\citenamefont{Gupta et~al.}(2015)\citenamefont{Gupta, Jager, Kats,
  Perez, and Stamou}}]{Gupta:2015zzs}
\bibinfo{author}{\bibfnamefont{R.~S.} \bibnamefont{Gupta}},
  \bibinfo{author}{\bibfnamefont{S.}~\bibnamefont{Jager}},
  \bibinfo{author}{\bibfnamefont{Y.}~\bibnamefont{Kats}},
  \bibinfo{author}{\bibfnamefont{G.}~\bibnamefont{Perez}}, \bibnamefont{and}
  \bibinfo{author}{\bibfnamefont{E.}~\bibnamefont{Stamou}}
  (\bibinfo{year}{2015}), \eprint{1512.05332}.

\bibitem[{\citenamefont{Petersson and Torre}(2016)}]{Petersson:2015mkr}
\bibinfo{author}{\bibfnamefont{C.}~\bibnamefont{Petersson}} \bibnamefont{and}
  \bibinfo{author}{\bibfnamefont{R.}~\bibnamefont{Torre}},
  \bibinfo{journal}{Phys. Rev. Lett.} \textbf{\bibinfo{volume}{116}},
  \bibinfo{pages}{151804} (\bibinfo{year}{2016}), \eprint{1512.05333}.

\bibitem[{\citenamefont{Molinaro et~al.}(2015)\citenamefont{Molinaro, Sannino,
  and Vignaroli}}]{Molinaro:2015cwg}
\bibinfo{author}{\bibfnamefont{E.}~\bibnamefont{Molinaro}},
  \bibinfo{author}{\bibfnamefont{F.}~\bibnamefont{Sannino}}, \bibnamefont{and}
  \bibinfo{author}{\bibfnamefont{N.}~\bibnamefont{Vignaroli}}
  (\bibinfo{year}{2015}), \eprint{1512.05334}.

\bibitem[{\citenamefont{Dutta et~al.}(2016)\citenamefont{Dutta, Gao, Ghosh,
  Gogoladze, and Li}}]{Dutta:2015wqh}
\bibinfo{author}{\bibfnamefont{B.}~\bibnamefont{Dutta}},
  \bibinfo{author}{\bibfnamefont{Y.}~\bibnamefont{Gao}},
  \bibinfo{author}{\bibfnamefont{T.}~\bibnamefont{Ghosh}},
  \bibinfo{author}{\bibfnamefont{I.}~\bibnamefont{Gogoladze}},
  \bibnamefont{and} \bibinfo{author}{\bibfnamefont{T.}~\bibnamefont{Li}},
  \bibinfo{journal}{Phys. Rev.} \textbf{\bibinfo{volume}{D93}},
  \bibinfo{pages}{055032} (\bibinfo{year}{2016}), \eprint{1512.05439}.

\bibitem[{\citenamefont{Cao et~al.}(2015)\citenamefont{Cao, Liu, Xie, Yan, and
  Zhang}}]{Cao:2015pto}
\bibinfo{author}{\bibfnamefont{Q.-H.} \bibnamefont{Cao}},
  \bibinfo{author}{\bibfnamefont{Y.}~\bibnamefont{Liu}},
  \bibinfo{author}{\bibfnamefont{K.-P.} \bibnamefont{Xie}},
  \bibinfo{author}{\bibfnamefont{B.}~\bibnamefont{Yan}}, \bibnamefont{and}
  \bibinfo{author}{\bibfnamefont{D.-M.} \bibnamefont{Zhang}}
  (\bibinfo{year}{2015}), \eprint{1512.05542}.

\bibitem[{\citenamefont{McAllister et~al.}(2016)\citenamefont{McAllister,
  Parker, and Tobar}}]{McAllister:2015zcz}
\bibinfo{author}{\bibfnamefont{B.~T.} \bibnamefont{McAllister}},
  \bibinfo{author}{\bibfnamefont{S.~R.} \bibnamefont{Parker}},
  \bibnamefont{and} \bibinfo{author}{\bibfnamefont{M.~E.} \bibnamefont{Tobar}},
  \bibinfo{journal}{Phys. Rev. Lett.} \textbf{\bibinfo{volume}{116}},
  \bibinfo{pages}{161804} (\bibinfo{year}{2016}), \eprint{1512.05547}.

\bibitem[{\citenamefont{Matsuzaki and Yamawaki}(2015)}]{Matsuzaki:2015che}
\bibinfo{author}{\bibfnamefont{S.}~\bibnamefont{Matsuzaki}} \bibnamefont{and}
  \bibinfo{author}{\bibfnamefont{K.}~\bibnamefont{Yamawaki}}
  (\bibinfo{year}{2015}), \eprint{1512.05564}.

\bibitem[{\citenamefont{Kobakhidze et~al.}(2016)\citenamefont{Kobakhidze, Wang,
  Wu, Yang, and Zhang}}]{Kobakhidze:2015ldh}
\bibinfo{author}{\bibfnamefont{A.}~\bibnamefont{Kobakhidze}},
  \bibinfo{author}{\bibfnamefont{F.}~\bibnamefont{Wang}},
  \bibinfo{author}{\bibfnamefont{L.}~\bibnamefont{Wu}},
  \bibinfo{author}{\bibfnamefont{J.~M.} \bibnamefont{Yang}}, \bibnamefont{and}
  \bibinfo{author}{\bibfnamefont{M.}~\bibnamefont{Zhang}},
  \bibinfo{journal}{Phys. Lett.} \textbf{\bibinfo{volume}{B757}},
  \bibinfo{pages}{92} (\bibinfo{year}{2016}), \eprint{1512.05585}.

\bibitem[{\citenamefont{Martinez et~al.}(2015)\citenamefont{Martinez, Ochoa,
  and Sierra}}]{Martinez:2015kmn}
\bibinfo{author}{\bibfnamefont{R.}~\bibnamefont{Martinez}},
  \bibinfo{author}{\bibfnamefont{F.}~\bibnamefont{Ochoa}}, \bibnamefont{and}
  \bibinfo{author}{\bibfnamefont{C.~F.} \bibnamefont{Sierra}}
  (\bibinfo{year}{2015}), \eprint{1512.05617}.

\bibitem[{\citenamefont{Cox et~al.}(2015)\citenamefont{Cox, Medina, Ray, and
  Spray}}]{Cox:2015ckc}
\bibinfo{author}{\bibfnamefont{P.}~\bibnamefont{Cox}},
  \bibinfo{author}{\bibfnamefont{A.~D.} \bibnamefont{Medina}},
  \bibinfo{author}{\bibfnamefont{T.~S.} \bibnamefont{Ray}}, \bibnamefont{and}
  \bibinfo{author}{\bibfnamefont{A.}~\bibnamefont{Spray}}
  (\bibinfo{year}{2015}), \eprint{1512.05618}.

\bibitem[{\citenamefont{Becirevic et~al.}(2016)\citenamefont{Becirevic,
  Bertuzzo, Sumensari, and Zukanovich~Funchal}}]{Becirevic:2015fmu}
\bibinfo{author}{\bibfnamefont{D.}~\bibnamefont{Becirevic}},
  \bibinfo{author}{\bibfnamefont{E.}~\bibnamefont{Bertuzzo}},
  \bibinfo{author}{\bibfnamefont{O.}~\bibnamefont{Sumensari}},
  \bibnamefont{and}
  \bibinfo{author}{\bibfnamefont{R.}~\bibnamefont{Zukanovich~Funchal}},
  \bibinfo{journal}{Phys. Lett.} \textbf{\bibinfo{volume}{B757}},
  \bibinfo{pages}{261} (\bibinfo{year}{2016}), \eprint{1512.05623}.

\bibitem[{\citenamefont{No et~al.}(2016)\citenamefont{No, Sanz, and
  Setford}}]{No:2015bsn}
\bibinfo{author}{\bibfnamefont{J.~M.} \bibnamefont{No}},
  \bibinfo{author}{\bibfnamefont{V.}~\bibnamefont{Sanz}}, \bibnamefont{and}
  \bibinfo{author}{\bibfnamefont{J.}~\bibnamefont{Setford}},
  \bibinfo{journal}{Phys. Rev.} \textbf{\bibinfo{volume}{D93}},
  \bibinfo{pages}{095010} (\bibinfo{year}{2016}), \eprint{1512.05700}.

\bibitem[{\citenamefont{Demidov and Gorbunov}(2016)}]{Demidov:2015zqn}
\bibinfo{author}{\bibfnamefont{S.~V.} \bibnamefont{Demidov}} \bibnamefont{and}
  \bibinfo{author}{\bibfnamefont{D.~S.} \bibnamefont{Gorbunov}},
  \bibinfo{journal}{JETP Lett.} \textbf{\bibinfo{volume}{103}},
  \bibinfo{pages}{219} (\bibinfo{year}{2016}), \eprint{1512.05723}.

\bibitem[{\citenamefont{Chao et~al.}(2015)\citenamefont{Chao, Huo, and
  Yu}}]{Chao:2015ttq}
\bibinfo{author}{\bibfnamefont{W.}~\bibnamefont{Chao}},
  \bibinfo{author}{\bibfnamefont{R.}~\bibnamefont{Huo}}, \bibnamefont{and}
  \bibinfo{author}{\bibfnamefont{J.-H.} \bibnamefont{Yu}}
  (\bibinfo{year}{2015}), \eprint{1512.05738}.

\bibitem[{\citenamefont{Fichet et~al.}(2016)\citenamefont{Fichet, von
  Gersdorff, and Royon}}]{Fichet:2015vvy}
\bibinfo{author}{\bibfnamefont{S.}~\bibnamefont{Fichet}},
  \bibinfo{author}{\bibfnamefont{G.}~\bibnamefont{von Gersdorff}},
  \bibnamefont{and} \bibinfo{author}{\bibfnamefont{C.}~\bibnamefont{Royon}},
  \bibinfo{journal}{Phys. Rev.} \textbf{\bibinfo{volume}{D93}},
  \bibinfo{pages}{075031} (\bibinfo{year}{2016}), \eprint{1512.05751}.

\bibitem[{\citenamefont{Curtin and Verhaaren}(2016)}]{Curtin:2015jcv}
\bibinfo{author}{\bibfnamefont{D.}~\bibnamefont{Curtin}} \bibnamefont{and}
  \bibinfo{author}{\bibfnamefont{C.~B.} \bibnamefont{Verhaaren}},
  \bibinfo{journal}{Phys. Rev.} \textbf{\bibinfo{volume}{D93}},
  \bibinfo{pages}{055011} (\bibinfo{year}{2016}), \eprint{1512.05753}.

\bibitem[{\citenamefont{Bian et~al.}(2016)\citenamefont{Bian, Chen, Liu, and
  Shu}}]{Bian:2015kjt}
\bibinfo{author}{\bibfnamefont{L.}~\bibnamefont{Bian}},
  \bibinfo{author}{\bibfnamefont{N.}~\bibnamefont{Chen}},
  \bibinfo{author}{\bibfnamefont{D.}~\bibnamefont{Liu}}, \bibnamefont{and}
  \bibinfo{author}{\bibfnamefont{J.}~\bibnamefont{Shu}},
  \bibinfo{journal}{Phys. Rev.} \textbf{\bibinfo{volume}{D93}},
  \bibinfo{pages}{095011} (\bibinfo{year}{2016}), \eprint{1512.05759}.

\bibitem[{\citenamefont{Chakrabortty et~al.}(2015)\citenamefont{Chakrabortty,
  Choudhury, Ghosh, Mondal, and Srivastava}}]{Chakrabortty:2015hff}
\bibinfo{author}{\bibfnamefont{J.}~\bibnamefont{Chakrabortty}},
  \bibinfo{author}{\bibfnamefont{A.}~\bibnamefont{Choudhury}},
  \bibinfo{author}{\bibfnamefont{P.}~\bibnamefont{Ghosh}},
  \bibinfo{author}{\bibfnamefont{S.}~\bibnamefont{Mondal}}, \bibnamefont{and}
  \bibinfo{author}{\bibfnamefont{T.}~\bibnamefont{Srivastava}}
  (\bibinfo{year}{2015}), \eprint{1512.05767}.

\bibitem[{\citenamefont{Ahmed et~al.}(2015)\citenamefont{Ahmed, Dillon,
  Grzadkowski, Gunion, and Jiang}}]{Ahmed:2015uqt}
\bibinfo{author}{\bibfnamefont{A.}~\bibnamefont{Ahmed}},
  \bibinfo{author}{\bibfnamefont{B.~M.} \bibnamefont{Dillon}},
  \bibinfo{author}{\bibfnamefont{B.}~\bibnamefont{Grzadkowski}},
  \bibinfo{author}{\bibfnamefont{J.~F.} \bibnamefont{Gunion}},
  \bibnamefont{and} \bibinfo{author}{\bibfnamefont{Y.}~\bibnamefont{Jiang}}
  (\bibinfo{year}{2015}), \eprint{1512.05771}.

\bibitem[{\citenamefont{Agrawal et~al.}(2015)\citenamefont{Agrawal, Fan,
  Heidenreich, Reece, and Strassler}}]{Agrawal:2015dbf}
\bibinfo{author}{\bibfnamefont{P.}~\bibnamefont{Agrawal}},
  \bibinfo{author}{\bibfnamefont{J.}~\bibnamefont{Fan}},
  \bibinfo{author}{\bibfnamefont{B.}~\bibnamefont{Heidenreich}},
  \bibinfo{author}{\bibfnamefont{M.}~\bibnamefont{Reece}}, \bibnamefont{and}
  \bibinfo{author}{\bibfnamefont{M.}~\bibnamefont{Strassler}}
  (\bibinfo{year}{2015}), \eprint{1512.05775}.

\bibitem[{\citenamefont{Csaki et~al.}(2016)\citenamefont{Csaki, Hubisz, and
  Terning}}]{Csaki:2015vek}
\bibinfo{author}{\bibfnamefont{C.}~\bibnamefont{Csaki}},
  \bibinfo{author}{\bibfnamefont{J.}~\bibnamefont{Hubisz}}, \bibnamefont{and}
  \bibinfo{author}{\bibfnamefont{J.}~\bibnamefont{Terning}},
  \bibinfo{journal}{Phys. Rev.} \textbf{\bibinfo{volume}{D93}},
  \bibinfo{pages}{035002} (\bibinfo{year}{2016}), \eprint{1512.05776}.

\bibitem[{\citenamefont{Aloni et~al.}(2015)\citenamefont{Aloni, Blum, Dery,
  Efrati, and Nir}}]{Aloni:2015mxa}
\bibinfo{author}{\bibfnamefont{D.}~\bibnamefont{Aloni}},
  \bibinfo{author}{\bibfnamefont{K.}~\bibnamefont{Blum}},
  \bibinfo{author}{\bibfnamefont{A.}~\bibnamefont{Dery}},
  \bibinfo{author}{\bibfnamefont{A.}~\bibnamefont{Efrati}}, \bibnamefont{and}
  \bibinfo{author}{\bibfnamefont{Y.}~\bibnamefont{Nir}} (\bibinfo{year}{2015}),
  \eprint{1512.05778}.

\bibitem[{\citenamefont{Gabrielli et~al.}(2016)\citenamefont{Gabrielli,
  Kannike, Mele, Raidal, Spethmann, and Veermae}}]{Gabrielli:2015dhk}
\bibinfo{author}{\bibfnamefont{E.}~\bibnamefont{Gabrielli}},
  \bibinfo{author}{\bibfnamefont{K.}~\bibnamefont{Kannike}},
  \bibinfo{author}{\bibfnamefont{B.}~\bibnamefont{Mele}},
  \bibinfo{author}{\bibfnamefont{M.}~\bibnamefont{Raidal}},
  \bibinfo{author}{\bibfnamefont{C.}~\bibnamefont{Spethmann}},
  \bibnamefont{and} \bibinfo{author}{\bibfnamefont{H.}~\bibnamefont{Veermae}},
  \bibinfo{journal}{Phys. Lett.} \textbf{\bibinfo{volume}{B756}},
  \bibinfo{pages}{36} (\bibinfo{year}{2016}), \eprint{1512.05961}.

\bibitem[{\citenamefont{Benbrik et~al.}(2016)\citenamefont{Benbrik, Chen, and
  Nomura}}]{Benbrik:2015fyz}
\bibinfo{author}{\bibfnamefont{R.}~\bibnamefont{Benbrik}},
  \bibinfo{author}{\bibfnamefont{C.-H.} \bibnamefont{Chen}}, \bibnamefont{and}
  \bibinfo{author}{\bibfnamefont{T.}~\bibnamefont{Nomura}},
  \bibinfo{journal}{Phys. Rev.} \textbf{\bibinfo{volume}{D93}},
  \bibinfo{pages}{055034} (\bibinfo{year}{2016}), \eprint{1512.06028}.

\bibitem[{\citenamefont{Alves et~al.}(2016)\citenamefont{Alves, Dias, and
  Sinha}}]{Alves:2015jgx}
\bibinfo{author}{\bibfnamefont{A.}~\bibnamefont{Alves}},
  \bibinfo{author}{\bibfnamefont{A.~G.} \bibnamefont{Dias}}, \bibnamefont{and}
  \bibinfo{author}{\bibfnamefont{K.}~\bibnamefont{Sinha}},
  \bibinfo{journal}{Phys. Lett.} \textbf{\bibinfo{volume}{B757}},
  \bibinfo{pages}{39} (\bibinfo{year}{2016}), \eprint{1512.06091}.

\bibitem[{\citenamefont{Carpenter et~al.}(2015)\citenamefont{Carpenter,
  Colburn, and Goodman}}]{Carpenter:2015ucu}
\bibinfo{author}{\bibfnamefont{L.~M.} \bibnamefont{Carpenter}},
  \bibinfo{author}{\bibfnamefont{R.}~\bibnamefont{Colburn}}, \bibnamefont{and}
  \bibinfo{author}{\bibfnamefont{J.}~\bibnamefont{Goodman}}
  (\bibinfo{year}{2015}), \eprint{1512.06107}.

\bibitem[{\citenamefont{Bernon and Smith}(2016)}]{Bernon:2015abk}
\bibinfo{author}{\bibfnamefont{J.}~\bibnamefont{Bernon}} \bibnamefont{and}
  \bibinfo{author}{\bibfnamefont{C.}~\bibnamefont{Smith}},
  \bibinfo{journal}{Phys. Lett.} \textbf{\bibinfo{volume}{B757}},
  \bibinfo{pages}{148} (\bibinfo{year}{2016}), \eprint{1512.06113}.

\bibitem[{\citenamefont{Khachatryan et~al.}(2015)}]{Khachatryan:2015qba}
\bibinfo{author}{\bibfnamefont{V.}~\bibnamefont{Khachatryan}}
  \bibnamefont{et~al.} (\bibinfo{collaboration}{CMS}), \bibinfo{journal}{Phys.
  Lett.} \textbf{\bibinfo{volume}{B750}}, \bibinfo{pages}{494}
  (\bibinfo{year}{2015}), \eprint{1506.02301}.

\bibitem[{\citenamefont{Huang et~al.}(2016)\citenamefont{Huang, Gu, Yin, Yu,
  and Zhang}}]{Huang:2015izx}
\bibinfo{author}{\bibfnamefont{F.~P.} \bibnamefont{Huang}},
  \bibinfo{author}{\bibfnamefont{P.-H.} \bibnamefont{Gu}},
  \bibinfo{author}{\bibfnamefont{P.-F.} \bibnamefont{Yin}},
  \bibinfo{author}{\bibfnamefont{Z.-H.} \bibnamefont{Yu}}, \bibnamefont{and}
  \bibinfo{author}{\bibfnamefont{X.}~\bibnamefont{Zhang}},
  \bibinfo{journal}{Phys. Rev.} \textbf{\bibinfo{volume}{D93}},
  \bibinfo{pages}{103515} (\bibinfo{year}{2016}), \eprint{1511.03969}.

\bibitem[{\citenamefont{Weinberg}(2013)}]{Weinberg:2013kea}
\bibinfo{author}{\bibfnamefont{S.}~\bibnamefont{Weinberg}},
  \bibinfo{journal}{Phys. Rev. Lett.} \textbf{\bibinfo{volume}{110}},
  \bibinfo{pages}{241301} (\bibinfo{year}{2013}), \eprint{1305.1971}.

\bibitem[{\citenamefont{Huang et~al.}(2014)\citenamefont{Huang, Li, Shao, and
  Wang}}]{Huang:2013oua}
\bibinfo{author}{\bibfnamefont{F.~P.} \bibnamefont{Huang}},
  \bibinfo{author}{\bibfnamefont{C.~S.} \bibnamefont{Li}},
  \bibinfo{author}{\bibfnamefont{D.~Y.} \bibnamefont{Shao}}, \bibnamefont{and}
  \bibinfo{author}{\bibfnamefont{J.}~\bibnamefont{Wang}},
  \bibinfo{journal}{Eur.Phys.J.} \textbf{\bibinfo{volume}{C74}},
  \bibinfo{pages}{2990} (\bibinfo{year}{2014}), \eprint{1307.7458}.

\bibitem[{\citenamefont{Zhan et~al.}(2016)\citenamefont{Zhan, Li, Liu, and
  Li}}]{Zhan:2015dha}
\bibinfo{author}{\bibfnamefont{Y.~C.} \bibnamefont{Zhan}},
  \bibinfo{author}{\bibfnamefont{C.~S.} \bibnamefont{Li}},
  \bibinfo{author}{\bibfnamefont{Z.~L.} \bibnamefont{Liu}}, \bibnamefont{and}
  \bibinfo{author}{\bibfnamefont{S.~A.} \bibnamefont{Li}},
  \bibinfo{journal}{Phys. Rev.} \textbf{\bibinfo{volume}{D93}},
  \bibinfo{pages}{014018} (\bibinfo{year}{2016}), \eprint{1508.02288}.

\bibitem[{\citenamefont{Martin et~al.}(2009)\citenamefont{Martin, Stirling,
  Thorne, and Watt}}]{Martin:2009iq}
\bibinfo{author}{\bibfnamefont{A.~D.} \bibnamefont{Martin}},
  \bibinfo{author}{\bibfnamefont{W.~J.} \bibnamefont{Stirling}},
  \bibinfo{author}{\bibfnamefont{R.~S.} \bibnamefont{Thorne}},
  \bibnamefont{and} \bibinfo{author}{\bibfnamefont{G.}~\bibnamefont{Watt}},
  \bibinfo{journal}{Eur. Phys. J.} \textbf{\bibinfo{volume}{C63}},
  \bibinfo{pages}{189} (\bibinfo{year}{2009}), \eprint{0901.0002}.

\bibitem[{CMS(2015{\natexlab{b}})}]{CMS-PAS-EXO-15-001}
\bibinfo{type}{Tech. Rep.} \bibinfo{number}{CMS-PAS-EXO-15-001},
  \bibinfo{institution}{The CMS Collaboration}
  (\bibinfo{year}{2015}{\natexlab{b}}),
  \urlprefix\url{http://cds.cern.ch/record/2048099}.

\bibitem[{\citenamefont{Alwall et~al.}(2014)\citenamefont{Alwall, Frederix,
  Frixione, Hirschi, Maltoni, Mattelaer, Shao, Stelzer, Torrielli, and
  Zaro}}]{Alwall:2014hca}
\bibinfo{author}{\bibfnamefont{J.}~\bibnamefont{Alwall}},
  \bibinfo{author}{\bibfnamefont{R.}~\bibnamefont{Frederix}},
  \bibinfo{author}{\bibfnamefont{S.}~\bibnamefont{Frixione}},
  \bibinfo{author}{\bibfnamefont{V.}~\bibnamefont{Hirschi}},
  \bibinfo{author}{\bibfnamefont{F.}~\bibnamefont{Maltoni}},
  \bibinfo{author}{\bibfnamefont{O.}~\bibnamefont{Mattelaer}},
  \bibinfo{author}{\bibfnamefont{H.~S.} \bibnamefont{Shao}},
  \bibinfo{author}{\bibfnamefont{T.}~\bibnamefont{Stelzer}},
  \bibinfo{author}{\bibfnamefont{P.}~\bibnamefont{Torrielli}},
  \bibnamefont{and} \bibinfo{author}{\bibfnamefont{M.}~\bibnamefont{Zaro}},
  \bibinfo{journal}{JHEP} \textbf{\bibinfo{volume}{07}}, \bibinfo{pages}{079}
  (\bibinfo{year}{2014}), \eprint{1405.0301}.

\bibitem[{\citenamefont{Aad et~al.}(2016{\natexlab{a}})}]{Aad:2016wpd}
\bibinfo{author}{\bibfnamefont{G.}~\bibnamefont{Aad}} \bibnamefont{et~al.}
  (\bibinfo{collaboration}{ATLAS}) (\bibinfo{year}{2016}{\natexlab{a}}),
  \eprint{1603.01702}.

\bibitem[{ATL(2013)}]{ATLAS-CONF-2013-020}
\bibinfo{type}{Tech. Rep.} \bibinfo{number}{ATLAS-CONF-2013-020},
  \bibinfo{institution}{The ATLAS Collaboration}, \bibinfo{address}{Geneva}
  (\bibinfo{year}{2013}), \urlprefix\url{http://cds.cern.ch/record/1525555}.

\bibitem[{\citenamefont{Aad et~al.}(2016{\natexlab{b}})}]{Aad:2015zqe}
\bibinfo{author}{\bibfnamefont{G.}~\bibnamefont{Aad}} \bibnamefont{et~al.}
  (\bibinfo{collaboration}{ATLAS}), \bibinfo{journal}{Phys. Rev. Lett.}
  \textbf{\bibinfo{volume}{116}}, \bibinfo{pages}{101801}
  (\bibinfo{year}{2016}{\natexlab{b}}), \eprint{1512.05314}.

\bibitem[{\citenamefont{Aad et~al.}(2016{\natexlab{c}})}]{Aad:2016sau}
\bibinfo{author}{\bibfnamefont{G.}~\bibnamefont{Aad}} \bibnamefont{et~al.}
  (\bibinfo{collaboration}{ATLAS}) (\bibinfo{year}{2016}{\natexlab{c}}),
  \eprint{1604.05232}.

\bibitem[{\citenamefont{Jungman et~al.}(1996)\citenamefont{Jungman,
  Kamionkowski, and Griest}}]{Jungman:1995df}
\bibinfo{author}{\bibfnamefont{G.}~\bibnamefont{Jungman}},
  \bibinfo{author}{\bibfnamefont{M.}~\bibnamefont{Kamionkowski}},
  \bibnamefont{and} \bibinfo{author}{\bibfnamefont{K.}~\bibnamefont{Griest}},
  \bibinfo{journal}{Phys. Rept.} \textbf{\bibinfo{volume}{267}},
  \bibinfo{pages}{195} (\bibinfo{year}{1996}), \eprint{hep-ph/9506380}.

\bibitem[{\citenamefont{Ade et~al.}(2015)}]{Ade:2015xua}
\bibinfo{author}{\bibfnamefont{P.~A.~R.} \bibnamefont{Ade}}
  \bibnamefont{et~al.} (\bibinfo{collaboration}{Planck})
  (\bibinfo{year}{2015}), \eprint{1502.01589}.

\bibitem[{\citenamefont{Alloul et~al.}(2014)\citenamefont{Alloul, Christensen,
  Degrande, Duhr, and Fuks}}]{Alloul:2013bka}
\bibinfo{author}{\bibfnamefont{A.}~\bibnamefont{Alloul}},
  \bibinfo{author}{\bibfnamefont{N.~D.} \bibnamefont{Christensen}},
  \bibinfo{author}{\bibfnamefont{C.}~\bibnamefont{Degrande}},
  \bibinfo{author}{\bibfnamefont{C.}~\bibnamefont{Duhr}}, \bibnamefont{and}
  \bibinfo{author}{\bibfnamefont{B.}~\bibnamefont{Fuks}},
  \bibinfo{journal}{Comput. Phys. Commun.} \textbf{\bibinfo{volume}{185}},
  \bibinfo{pages}{2250} (\bibinfo{year}{2014}), \eprint{1310.1921}.

\bibitem[{\citenamefont{Sjostrand et~al.}(2006)\citenamefont{Sjostrand, Mrenna,
  and Skands}}]{Sjostrand:2006za}
\bibinfo{author}{\bibfnamefont{T.}~\bibnamefont{Sjostrand}},
  \bibinfo{author}{\bibfnamefont{S.}~\bibnamefont{Mrenna}}, \bibnamefont{and}
  \bibinfo{author}{\bibfnamefont{P.~Z.} \bibnamefont{Skands}},
  \bibinfo{journal}{JHEP} \textbf{\bibinfo{volume}{05}}, \bibinfo{pages}{026}
  (\bibinfo{year}{2006}), \eprint{hep-ph/0603175}.

\bibitem[{\citenamefont{Butterworth et~al.}(2008)\citenamefont{Butterworth,
  Davison, Rubin, and Salam}}]{Butterworth:2008iy}
\bibinfo{author}{\bibfnamefont{J.~M.} \bibnamefont{Butterworth}},
  \bibinfo{author}{\bibfnamefont{A.~R.} \bibnamefont{Davison}},
  \bibinfo{author}{\bibfnamefont{M.}~\bibnamefont{Rubin}}, \bibnamefont{and}
  \bibinfo{author}{\bibfnamefont{G.~P.} \bibnamefont{Salam}},
  \bibinfo{journal}{Phys. Rev. Lett.} \textbf{\bibinfo{volume}{100}},
  \bibinfo{pages}{242001} (\bibinfo{year}{2008}), \eprint{0802.2470}.

\bibitem[{\citenamefont{Cacciari et~al.}(2012)\citenamefont{Cacciari, Salam,
  and Soyez}}]{Cacciari:2011ma}
\bibinfo{author}{\bibfnamefont{M.}~\bibnamefont{Cacciari}},
  \bibinfo{author}{\bibfnamefont{G.~P.} \bibnamefont{Salam}}, \bibnamefont{and}
  \bibinfo{author}{\bibfnamefont{G.}~\bibnamefont{Soyez}},
  \bibinfo{journal}{Eur. Phys. J.} \textbf{\bibinfo{volume}{C72}},
  \bibinfo{pages}{1896} (\bibinfo{year}{2012}), \eprint{1111.6097}.

\bibitem[{\citenamefont{collaboration}(2016)}]{ATLAS:2016eeo}
\bibinfo{author}{\bibfnamefont{T.~A.} \bibnamefont{collaboration}}
  (\bibinfo{collaboration}{ATLAS}) (\bibinfo{year}{2016}).

\bibitem[{\citenamefont{Collaboration}(2016)}]{CMS:2016crm}
\bibinfo{author}{\bibfnamefont{C.}~\bibnamefont{Collaboration}}
  (\bibinfo{collaboration}{CMS}) (\bibinfo{year}{2016}).

\end{thebibliography}

\end{document}